\documentclass[%
 reprint,
superscriptaddress,
 amsmath,amssymb,
 aps,
 prb,
]{revtex4-2}
\usepackage{graphicx}% Include figure files
\usepackage{dcolumn}% Align table columns on decimal point
\usepackage{bm}% bold math
\usepackage[version=4]{mhchem}
\usepackage{xcolor}
\usepackage{braket}
\begin{document}

\preprint{APS/123-QED}

\title{Exciton-carrier coupling in a metal halide perovskite nanocrystal assembly probed by two-dimensional coherent spectroscopy}
% Force line breaks with \\

\author{Esteban Rojas-Gatjens}%
\email[Electronic address: ]{esteban.rojas@gatech.edu}
\affiliation{School of Chemistry and Biochemistry, Georgia Institute of Technology, Atlanta, GA, United~States\\}%
\affiliation{School of Physics, Georgia Institute of Technology, Atlanta, GA, United~States\\}

\author{David Otto Tiede}
\affiliation{Institute of Materials Science of Sevilla, Spanish National Research Council, Seville, Spain\\}
\affiliation{Department of Physics and Center for Functional Materials, Wake Forest University, Winston-Salem, NC, United~States\\}

\author{Katherine A. Koch}
\affiliation{Department of Physics and Center for Functional Materials, Wake Forest University, Winston-Salem, NC, United~States\\}

\author{Carlos Romero-Perez}
\affiliation{Institute of Materials Science of Sevilla, Spanish National Research Council, Seville, Spain\\}

\author{Juan F. Galisteo-López}
\affiliation{Institute of Materials Science of Sevilla, Spanish National Research Council, Seville, Spain\\}

\author{Mauricio E. Calvo}
\affiliation{Institute of Materials Science of Sevilla, Spanish National Research Council, Seville, Spain\\}

\author{Hern{\'a}n M{\'i}guez}
\affiliation{Institute of Materials Science of Sevilla, Spanish National Research Council, Seville, Spain\\}

\author{Ajay~Ram~Srimath~Kandada}
\email[Electronic address: ]{srimatar@wfu.edu}
\affiliation{Department of Physics and Center for Functional Materials, Wake Forest University, Winston-Salem, NC, United~States\\}%

\date{\today}% It is always \today, today,

\begin{abstract}
The surface chemistry and inter-connectivity within perovskite nanocrystals play a critical role in determining the electronic interactions. They manifest in the Coulomb screening of electron-hole correlations and the carrier relaxation dynamics, among other many-body processes. Here, we characterize the coupling between the exciton and free carrier states close to the band-edge in a ligand-free formamidinium lead bromide nanocrystal assembly via two-dimensional coherent spectroscopy. The optical signatures observed in this work show: (i) a nonlinear spectral lineshape reminiscent of Fano-like interference that evidences the coupling between discrete electronic states and a continuum, (ii) symmetric excited state absorption cross-peaks that suggest the existence of a coupled exciton-carrier excited state, and (iii) ultrafast carrier thermalization and exciton formation. Our results highlight the presence of coherent coupling between exciton and free carriers, particularly in the sub-100 femtosecond timescales. %$\sim 100$\,fs time scale.

\end{abstract}

%\keywords{Perovskite nanocrystal, Two-dimensional electronic spectroscopy, Many-body interactions, Exciton-carrier coupling, hot-carrier relaxation}

\maketitle

\section{\label{sec:intro1}Introduction}

Lead halide perovskite nanocrystals (PNCs) are promising candidates for quantum dot optoelectronic applications~\cite{Abhishek2016, Chiba2018}. Recently, device-compatible assemblies of PNCs grown in a porous scaffold have been shown to exhibit exceptional optoelectronic properties~\cite{Dirin2016Harnessing, Malgras2016Observation, rubino2020mesoporous}. The absence of the organic ligands on the surface of the PNCs may however activate surface defect states which will, depending on the relative energetics, either act as sources for carrier doping~\cite{kandada2016} or centers for non-radiation recombination of the photo-exctiations~\cite{Jariwala2021}. On the other hand, their absence promotes effective inter-connectivity and thus electronic coupling between the PNCs, which subsequently leads to much-desired efficient charge and energy transport~\cite{Rubino2021, romero2022optoelectronic} within the PNC assembly. Notably, in comparison to the more widely studied colloidal systems, these PNC assemblies present a distinct photophysical scenario in which inter-particle interactions are expected to play a dominant role in photo-excitation dynamics. Unraveling these interactions is of utmost relevance for the further optimization of this alternative approach to PNC-based materials.      
% eventually affect the electronic properties by doping~\cite{kandada2016} or non-radiative recombination~\cite{Jariwala2021}. In addition, 
%In recent work, distinct groups have prepared PNC assemblies by growing them in a porous scaffold~\cite{Dirin2016Harnessing, Malgras2016Observation, rubino2020mesoporous}. 
%The surface of these nanocrystals is not covered by ligands, which implies that they could potentially interact provided the average distance between them is short enough. Also, potential surface defects might not be passivated. which may eventually affect the electronic properties by doping~\cite{kandada2016} or non-radiative recombination~\cite{Jariwala2021}
%Additionally, it has been observed that the absence of ligands results in enhanced charge transport and nanocrystal interconnectivity in nanocrystal assemblies, resulting in increased performances in devices like solar cells or photovoltaic devices~\cite{Rubino2021, romero2022optoelectronic}.
%Unraveling the exciton and carriers' interplay is therefore of utmost relevance due to the technological promise held by this alternative approach to PNC-based materials.

It is now well known that the photoluminescence in the PNCs originates from radiative recombination of bound electron-hole pairs (excitons). 
The correlation between excitons and free charge carriers  %
%The correlation of bound electron-hole pairs (excitons) and free carriers 
has been a topic of research for over two decades, with early works on semiconducting systems such as GaAs~\cite{Wehner1996, Webber2016} and InP~\cite{Allan1999}. Signatures of substantial exciton-carrier interactions have also been identified recently in bulk lead-halide perovskites ~\cite{Trinh2015, Jha2018, Palmieri2020Mahan, Nguyen2019}. The many-body correlations between excitons and carriers manifest in the nonlinear optical response of the materials, specifically in the evolution in the spectral lineshapes in the ultrafast timescales. % properties of the materials, however, through linear spectroscopy, these signatures are difficult to disentangle. Differential transmission spectroscopy offers insight into the many-body interactions of carriers by analyzing the differential lineshape and population evolution~\cite{Trinh2015, Wu2015}. 
Two-dimensional coherent spectroscopy is the technique of preference for disentangling such many-body correlations in bulk and nanostructured semiconductors~\cite{Li2006, Moody2011, Stone2009, Turner2010, Collini2020, Collini2021, Liu_2022}.
% since it provides energy resolution on the excitation axis and phase information. 
The correlations between excited states manifest as energy shifts, resonances of multi-particle states (e.g. biexcitons and trions), spectral linewidth and phase shifts in the complex lineshape of the two-dimensional excitation-emission map ($\hbar \omega_1$-$\hbar \omega_3$, respectively)~\cite{Li2006, Webber2016, Kandada2020Stochastic, Allerber2021, Rojas2023manyexciton}.
Several groups have used two-dimensional coherent spectroscopy to describe the charge carrier thermalization, population relaxation, and exciton dissociation in bulk lead-halide perovskite semiconductors~\cite{Richter2017, Jha2018, Nguyen2019}. 
%Perovskite nanocrystals are of particular interest since the spatial confinement and the surface chemistry heavily influence the confined exciton state and the free carrier continuum photophysics. 
In the context of PNCs, Yu~\textit{et al}~\cite{Yu2021} used two-dimensional spectroscopy to evaluate the bottleneck effect in the hot carrier thermalization as a function of nanocrystal size. Most of such investigations in semiconductor nanostructures have been performed on colloidal suspensions. Notably, exciton-carrier correlations have not been rigorously explored in such systems. Moreover, the colloidal systems lack the interconnectivity between the particles to explore multi-particle correlations and in fact, only a handful of works on solid-state assemblies have investigated coherent inter-particle interactions~\cite{Collini2020, Collini2021}.  % observing that the smaller the nanocrystal size the longer it would take the carriers to cool. 
%However, in general, for nanocrystals, most of the 2D coherent spectroscopic studies have been conducted in colloidal suspension, with a few exceptions in solid-state assemblies where the studies were interested in interparticle coupling~\cite{Collini2020, Collini2021}.    

Here, we investigate the exciton-carrier coupling and inter-particle interactions through 2D coherent spectroscopy~\cite{Turner2011} in solid dispersions of ligand-free formamidinium lead bromide (\ce{FAPbBr3}) nanocrystals embedded in a nanoporous silica scaffold. Through a systematic analysis of the observed spectral lineshape, we discuss the evidence for electronic coupling between band-edge carrier states and discrete excitonic transitions within an interconnected network of PNCs. We identify excited state absorption features associated with a trion-like coupled exciton-carrier state. In addition, we reproduce the observed experimental lineshapes using a photophysical model in which discrete exciton and band-edge carrier states interact with the carrier continuum via Fano-like interference mechanism. Lastly, we discuss the time evolution of the spectral response in the context of carrier relaxation into the excitonic state that happens within $\approx$ 100\,fs.  
%As described above, the preparation conditions result in a surface prone to defects but also with an enhanced interconnectivity between nanocrystals. 
%In this work, we present evidence, attained by two-dimensional coherent spectroscopy, of the coupling between band-edge states and discrete confined transitions that occurs in an interconnected network of perovskite quantum nanocrystals dispersed in a porous matrix. Signatures of such interaction are: (i) An asymmetric lineshape which we interpret as a Fano-like interference between the discrete confined state and a continuum of states, (ii) an excited-state absorption cross peak which we assigned to an exciton-carrier coupled state, and (iii) the appearance of a positive cross-peak emitted at lower energies which evidences thermalization from the free carriers to the excitons.

\begin{figure}
    \centering
    \includegraphics[width=\linewidth]{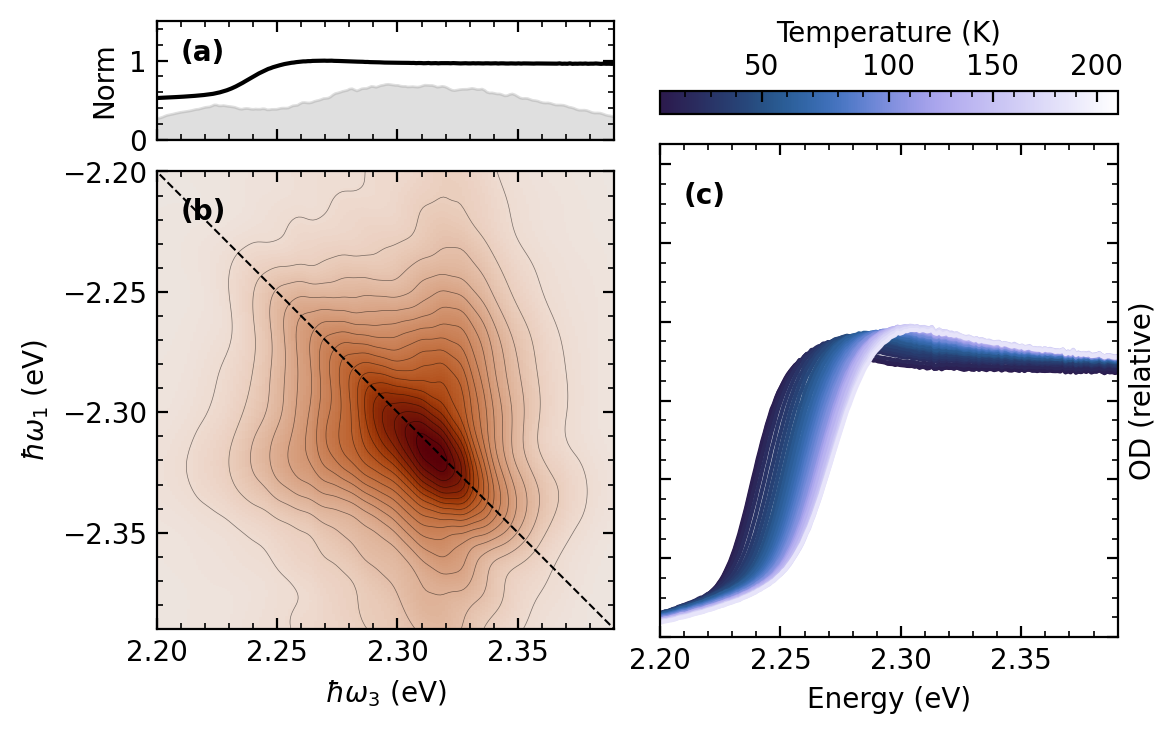}
    \caption{Absorption and rephasing (absolute) spectra for a \ce{FAPbBr3} Ncs thin film. (a) Absorption (black line) and laser spectrum used in the 2D coherent electronic spectroscopy (gray area). (b) Coherent 1Q-rephasing response, $\Vec{k}_{sig} = -\Vec{k}_{a}+\Vec{k}_{b}+\Vec{k}_{c}$, showing the absolute component of the spectrum. (c) Temperature dependence of the Absorption measurements for \ce{FAPbBr3} NCs thin films. (a) and (b) were measured at 10\,K.}
    \label{fig:Absolute}
\end{figure}

\section{\label{sec:Result}Results and Discussion}

The \ce{FAPbBr3} NCs solid-state assembly is prepared by infiltrating a precursor solution to the void space of the nanoporous silica scaffold via spin-coating followed by an annealing step. More details on the sample preparation can be found in the Supporting Information and in Ref.~\citenum{romero2023}. The \ce{FAPbBr3} nanocrystals self-assemble within the pores of the matrix, with the precursor concentration determining the average particle size and the filling fraction. % and aver resulting are \ce{FAPbBr3} nanocrystals within the pores of the matrix. 
For a nominal concentration of 30\% we obtained PNCs of the average diameter of 6.7\,nm and a pore fill fraction of 0.145, estimated through inductively coupled plasma analysis (details in the SI). Note that since the PNCs are crystallized in the voids of the porous matrix, the geometry, and average size can undergo variations that cannot be rigorously quantified. These samples show a photoluminescence quantum yield of approximately 5\%, moderately higher than what is observed in bulk films. Given that the Bohr radius in \ce{FAPbBr3} is estimated to be about 8\,nm~\cite{Perumal2016}, the PNC assembly used here can be considered to be in an intermediate confinement regime. Overall, the sample used in this study is a solid-state assembly of \emph{ligand-free} perovskite NCs with enhanced charge coupling between the distinct particle units compared to nanocrystals with ligands.  
%The estimated Bohr radius is 8\,nm~\cite{Perumal2016}, indicating that the nanocrystals are in an intermediate confinement regime. 
%This methodology results in a grid of \emph{ligand-free} perovskite NCs with enhanced charge coupling between the distinct particle units compared to nanocrystals with ligands.

The temperature-dependent absorption spectrum of the sample is shown in Fig.~\ref{fig:Absolute}(a). We observe a spectral lineshape typical of a direct bandgap semiconductor, associated with the optical absorption of a carrier continuum and a low binding energy exciton band close to the edge, at all temperatures. We observe a blueshift in the optical edge in the PNCs in comparison to the absorption spectrum of the bulk \ce{FAPbBr3} film (see Supporting Information), supporting the electronic confinement. Interestingly, the spectrum conspicuously lacks an evident excitonic peak, which is clearly present in the spectrum of the bulk film even at room temperature. There may be a number of causes for the reduction in the exciton binding energy in these PNC assemblies, including inter-particle coupling~\cite{Baumgardner2013} and Coulomb screening from background doping~\cite{Palmieri2020Mahan}. We will not discuss the origin of the exciton screening in this manuscript, but we highlight the reduced binding energy that results in spectrally close exciton and continuum bands. The spectral broadening either due to the polydispersity or large intrinsic dephasing rate results in substantial spectral overlap between the discrete exciton states and the continuum, which can promote effective interaction between the two species. We investigate such correlations with 2D coherent spectroscopy, which we briefly discuss in the following.       
%In the supplementary material, we compared the bulk \ce{FAPbBr3} absorption with the nanocrystal measurement at room temperature. We observe a blueshift for the case of the nanocrystal, evidence of electronic confinement. 
%Interestingly the excitonic peak is not as evident for the nanocrystals as it is for the bulk material at room temperature. This has been attributed as a consequence of polydispersity and inter-dot coupling~\cite{Baumgardner2013}. It could also be a result of a background doping effect which screens the exciton binding energy~\cite{Palmieri2020Mahan} or screening through the motion of free surface cations~\cite{even2014analysis}.

%Let us briefly describe the 2D coherent spectroscopy used in this work. 
The experiment consists of a train of three ultrashort pulses incident on the sample in a BoxCAR geometry. Such a coherent excitation sequence generates a third-order nonlinear polarization in the material, which emits coherent radiation in the phase-matching direction. A fourth pulse that serves as a local oscillator enables measurement of the amplitude and phase of the emitted field. 
The first pulse generates a coherence that evolves for a time ($t_1$), the second pulse further drives the system to a population state which relaxes within a population time ($t_2$), and finally, the third pulse establishes another coherence which emits the electric field.
We measure the emitted electric field from the ensemble of photo-excited species through spectral interferometry whose energy corresponds with the \textit{emission} axis of the two-dimensional spectrum, labeled $\hbar \omega_3$ here. By scanning $t_1$ and performing a discrete Fourier transform we recover the \textit{excitation axis} which we label as $\hbar\omega_1$. We specifically focus on the nonlinear response with a phase-matching condition, $\Vec{k}_{sig} = -\Vec{k}_{a}+\Vec{k}_{b}+\Vec{k}_{c}$, referred to as the \textit{rephasing} pathway. The experimental details are described in the supplementary information and further details can be found elsewhere~\cite{Turner2011}.

The spectrum of the ultrashort pulses used in the current experiment covers the exciton and free carrier energies close to the optical edge of the sample, as shown in %In the 2D coherent experiments presented, the ultrafast pulse excites the band edge of the material, simultaneously covering the exciton states and the free carriers band, as shown in 
Fig.~\ref{fig:Absolute}(a). We show the absolute value of the rephasing spectrum of the sample held at 10\,K and taken at a population time of 10\,fs in Fig.~\ref{fig:Absolute}(b). The extended feature along the diagonal line confirms the early population of the excited states in the experimental spectral range. The larger intensity at the higher energies indicates a larger fraction of the initial population in the free carrier states at early times. More importantly, we observe clear off-diagonal cross-peaks that indicate coherent coupling between exciton states and the free carriers. In the simplest scenario, this can be interpreted as a consequence of the same ground state shared by both the exciton and free carrier states.  

\begin{figure}
    \centering
    \includegraphics[width=\linewidth]{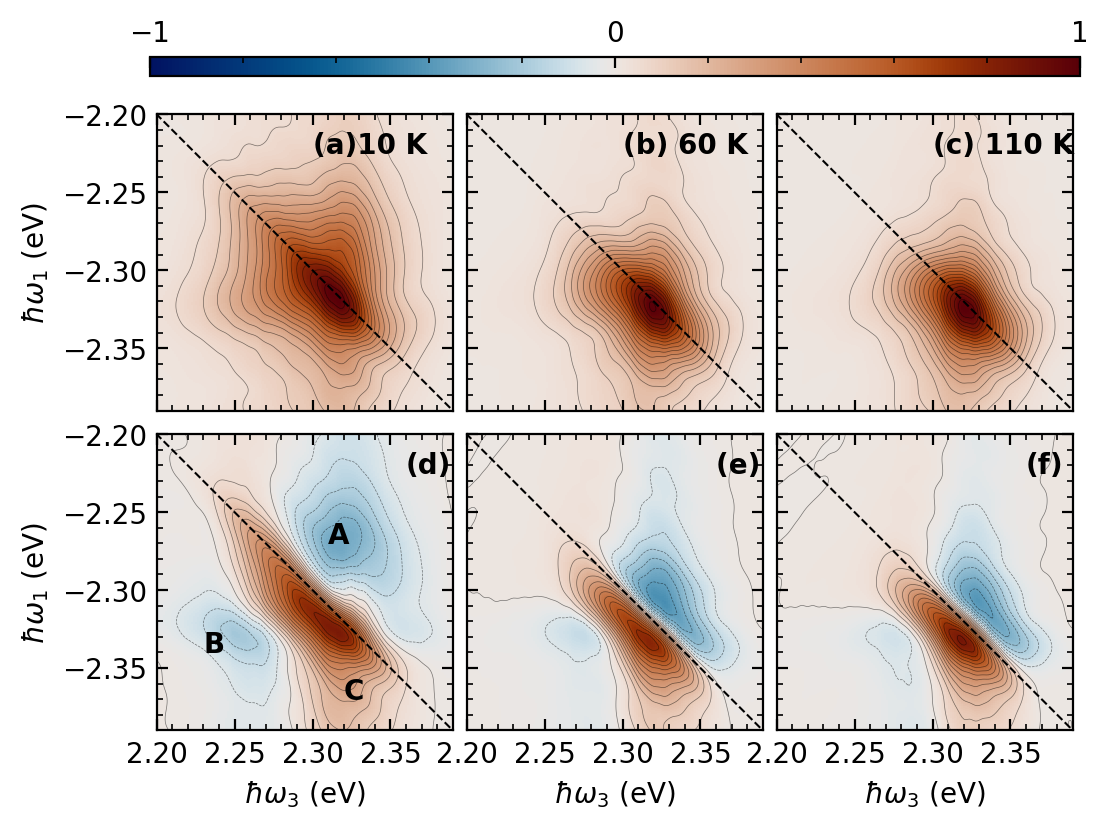}
    \caption{Absolute and real components of the 2D coherent spectrum at the top and bottom respectively. They were measured at temperatures (10 K, 60 K, and 110 K) for a \ce{FAPbBr3} NCs thin film. .}
    \label{fig:TempDep}
\end{figure}

More details on the excited-state couplings can be deduced by analyzing the real component of the rephasing spectrum along with the absolute spectrum at different temperatures as shown in Fig.~\ref{fig:TempDep}. In the real part of the response at 10\,K (Fig.~\ref{fig:TempDep}(d)), we observe that the cross-peaks correspond to negative excited-state absorption (ESA) features, labeled as \textbf{A} and \textbf{B}. In addition, we observe an extended positive off-diagonal \textit{streak} at higher energies labeled \textbf{C}. We interpret the latter as a signature of Fano-like interference which will be discussed below. %We will discuss how these optical signatures suggest two physical insights: (i) there is a coupled state composed of an exciton and a carrier and (ii) exciton shows Fano-like interference with a continuum of states.

Firstly, the ESA feature in a 2D spectrum is associated with an excitation pathway that results in the emitted field oscillating at the energy of the coherence between the photo-generated species and a higher-lying state. The latter is typically a two-quantum transition where the energy of the final state depends on the correlations between the excited species~\cite{Yang2007Correlations}. For example, two excitons with attractive interactions result in a biexciton state, whose energy is less than twice the energy of the exciton. Thus the ESA feature associated with the exciton to biexciton transition will appear as an off-diagonal negative peak and below the diagonal in a 2D spectrum~\cite{Yang2008Isolating}. 
%In the excited state absorption pathway, the third pulse further excites the population states to a coherent state with a higher-lying energy state. It can be interpreted as a two-quantum transition where the energy of the final state depends on the correlations between the excited species~\cite{Yang2007Correlations}, for example, two excitons with attractive interactions result in a biexciton state which will appear as an off-diagonal peak. The correlation between two transitions with different energies is expressed as cross-peaks~\cite{Yang2008Isolating}. We will now justify the assignment of excited state absorption features \textbf{A} and \textbf{B} to coupled states. 
However, we discard such biexcitonic transitions as the origin for the observed ESA features as they would not result in a symmetric absolute lineshape as is the case for our experimental data. 
While one could explain \textbf{B} as a biexcitonic state associated with the diagonal feature at 2.33\,eV, to explain \textbf{A} one would have to invoke a repulsive two-quantum state in addition to the attractive biexciton that makes it physically unfeasible. % that is unlikely to show high transition dipoles and lifetimes. As discussed below, the signature \textbf{A} leaves longer than 120\,fs, this argues against a biexciton with repulsive interactions.
Notably, previous works on PNCs~\cite{Huang2020, Zhao2019} that have identified biexcitonic features in a 2D spectrum had clear asymmetric spectral features, unlike the present case. % Other authors studying perovskite nanocrystals have observed ESA features and assigned them to biexcitonic transitions and the transition from an exciton to higher energy states~\cite{Huang2020, Zhao2019}.  
%However, in those reports, there is a clear asymmetric structure typical of attractive biexcitonic states.
Feature \textbf{A} is associated with a transition from the state populated at 2.25\,eV, at the exciton energy to a higher-lying state at an energy that is slightly higher than twice the exciton energy. Feature \textbf{B} instead originates from a populated state at $\approx$ 2.3\,eV and thus in the carrier continuum to a stable two-quantum state. Given the symmetry in the features, we can assign the features to transitions to the same two-quantum state, one from the exciton state and the other from the free carrier state. The relative energetics indicate that the two-quantum state is in fact a coupled exciton-carrier state.  
%The ESA pathway of coherently coupled states will result in symmetric negative cross-peaks. 
%We assign the lower energy states to exciton and, therefore, assign the \textbf{A} and \textbf{B} signatures to exciton-carrier coupled states.
%, exemplified in the double-Feynman diagram (ESA1, Fig~\ref{fig:pathways}).

\begin{figure}
    \centering
    \includegraphics[width=\linewidth]{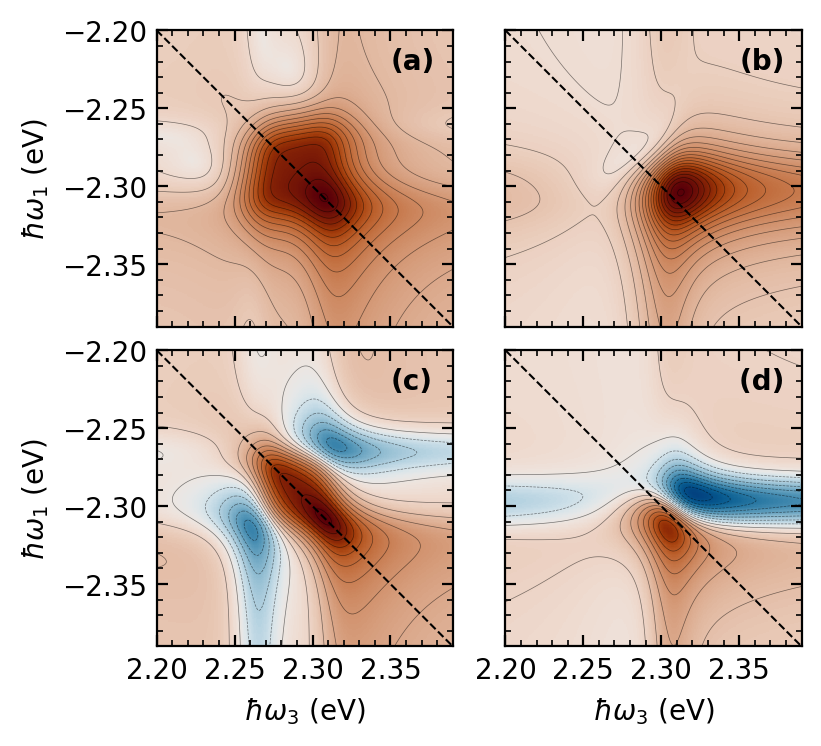}
    \caption{Absolute values of the simulated rephasing response for two discrete states separated by (a) 40\,meV and (b) 20\,meV, which includes contributions from coupled excited state pathways. The coupling with the continuum was set to $q=1$. The corresponding real parts of the response are shown in (c) and (d) respectively.}
    \label{fig:simul}
\end{figure}

\begin{figure*}[ht]
    \centering
    \includegraphics[width=1\linewidth]{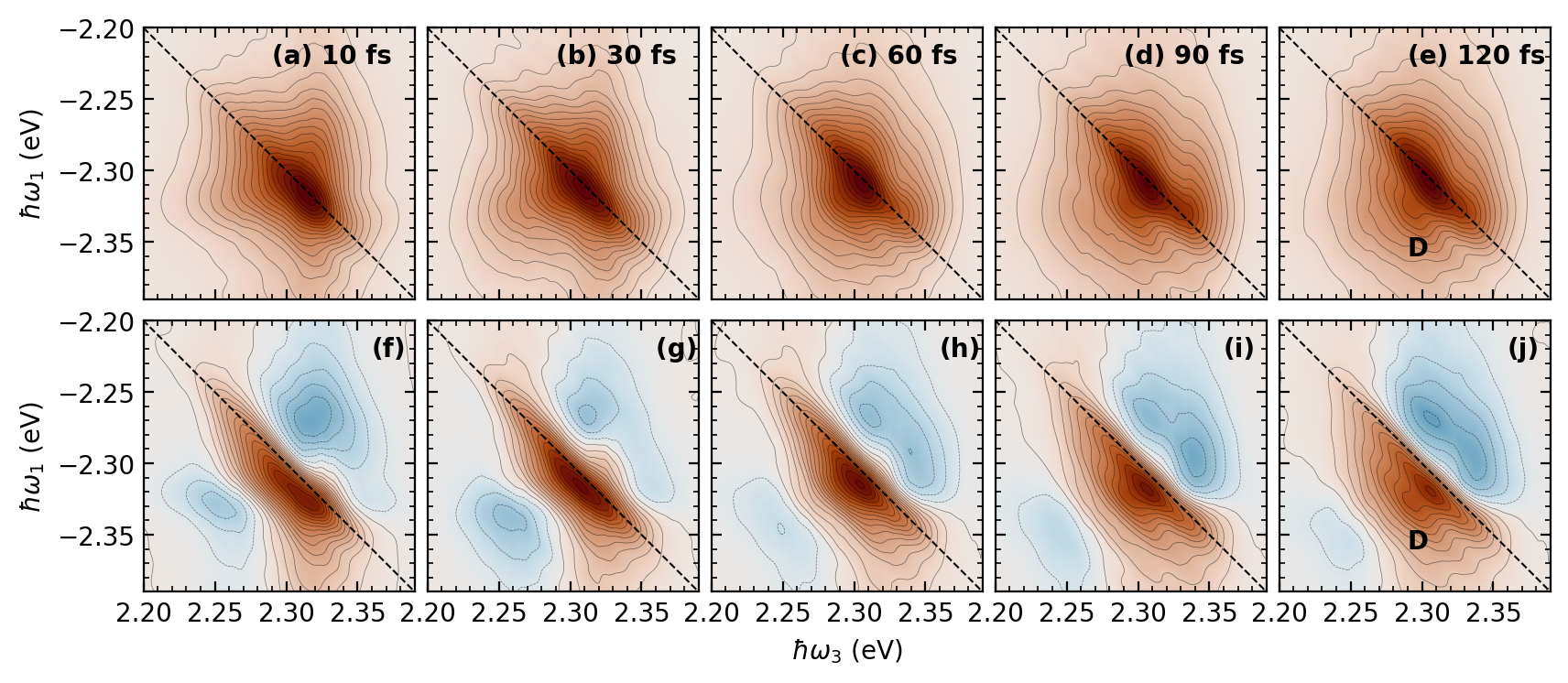}
    \caption{Evolution of the rephasing 2D coherent spectra as a function of population time. The subfigures (a-e) correspond to the absolute components and (f-j) correspond to the real component.}
    \label{fig:fig_pop} 
\end{figure*}

As the temperature increases the absorption edge shifts towards higher energy, as evident in the spectra shown in Figs~\ref{fig:Absolute}(b) and (c). Similarly, the 2D coherent spectrum blue shifts and the real component evolves from a symmetric \textit{absorptive} lineshape, Fig~\ref{fig:TempDep}(d), with very clear cross-peaks to a \textit{dispersive} lineshape, Fig~\ref{fig:TempDep}(e) and (f). Extensive literature describes the dispersive lineshapes observed in semiconductors in four-wave mixing experiments~\cite{Li2006, Kandada2020Stochastic}. The observed spectral asymmetry and derivative-like lineshape along the anti-diagonal of the 2D spectrum are typically attributed to many-body interactions which result in an excitation-induced shift and excitation-induced dephasing process~\cite{Li2006, Shacklette2002, Shacklette2003, Bristow2009, Li2020, Li2023, Trovatello2022}. In such a context the observed dispersive lineshapes at higher temperatures may imply temperature-induced many-body correlations between photo-excitations. It is possible that higher temperatures promote larger background doping, thus driving the observed lineshapes. However, we do not have independent confirmation of such a process. Acknowledging the need to further rationalize these lineshapes at higher temperatures, we interpret the observed dispersive-like lineshape as a consequence of the overlap of two spectral features. Specifically, the blue shift of the cross peaks that accompanies the shift of exciton's main feature and results in a spectral overlap of the diagonal and off-diagonal features, resulting in a seemingly dispersive line shape. This type of behavior due to overlapping ground state bleach (GB) and ESA features is commonly observed in 2D infrared spectroscopy for systems with strong anharmonicity~\cite{Baiz2020}. %Additionally, as the temperature increases the two states, distinguishable at low-temperature, overlap and resemble a single transition.

Having described the signatures of ESA, \textbf{A} and \textbf{B}, we now turn our attention to the extended cross-peak observed in all of the real spectra of Fig.~\ref{fig:TempDep}(d-f), and labeled \textbf{C} in Fig.~\ref{fig:TempDep}(d).
Similar extended features at high $\hbar\omega_1$ have been observed in the case of \ce{GaAs} bulk semiconductors~\cite{Webber2016, Wilmer2016} and arise due to correlations between the excitons and the free carrier continuum as they share a common ground state. Nguyen~\textit{et al}~\cite{Nguyen2019} observed similar features also in lead-bromide perovskite single crystals. In the case of GaAs ~\cite{Webber2016, Wilmer2016}, the exciton transitions are well-defined and energetically resolved from the free carrier transition, especially at low temperatures. In the present case, however, the distribution of exciton states and band-edge tail energetically overlap. Given such an overlap, a Fano-like interference between the discrete exciton state and the continuum state may be considered~\cite{Fano1961, Fano1}. Finkelstein-Shapiro \textit{et al.} developed an analytical model that describes such a scenario in which the discrete state's energy lies in the middle of the continuum distribution and predicted 2D lineshapes similar to what is observed here~\cite{Fano1, Fano2}. Extending this model, we consider here two discrete states representing the exciton and the free-carrier band edge interacting with a continuum of states. The continuum corresponds to the higher energy free carrier states but could also have distinct physical origins. For example, a continuum can be a result of the inhomogeneous distributions of coupled nanocrystals, Urbach tail states~\cite{Li2023_2}, disorder-induced energy dispersion~\cite{G2009}, among others~\cite{Tiguntseva2018}. Without dwelling on the origin and nature of the continuum, We apply the analytical expressions derived by Finkelstein-Shapiro~\textit{et al.}, summarized in supporting information, to simulate the expected nonlinear response, shown in Fig.~\ref{fig:simul}.  %With the data presented in this work, we cannot confidently assign a physical scenario.
%Finkelstein-Shapiro~\textit{et al.} derived the 2D coherent lineshape expected for a Fano-like model~\cite{Fano1, Fano2}. 

Firslty, we consider two main transitions associated with the exciton and free carrier band edge to be at energies $k_1 =2.27$\,eV and $k_2 = 2.31$\,eV respectively (energy separation of 40\,meV). The strength of the coupling of these states with the continuum is determined by the parameter $q$, $q = \frac{\mu_e}{\mu_c \pi V}$ where $V$ is the coupling constant between the continuum of states and the discrete transition. For the simulations showed here, we kept the value of $q$ to be 1. We also considered contributions from two excited state absorption transitions from $k_1$ or $k_2$ to a coupled state $k_1k_2$. More details on the analytical expressions used for the simulations are given the Supporting Information (Sec. 2.2). For this set of simulation parameters, we show the absolute and real values of the rephasing spectrum in Figs~\ref{fig:simul}(a) and (c) respectively. It can be seen that the simulation qualitatively reproduces the low temperature experimental spectra shown in Figs.\ref{fig:TempDep}(a) and (d). 

We now reduce the energy difference between the discrete exciton and free-carrier band edge state to 20\,meV and obtain rephasing response shown in Fig.~\ref{fig:simul}(b) and (d). It can be seen in Fig.~\ref{fig:simul}(d) that 2D lineshape acquires a dispersive-like spectral feature and qualitatively resembles the experimental lineshapes at higher temperatures (Fig.~\ref{fig:TempDep} (e) and (f)). This confirms our initial hypothesis that the dispersive lineshape here is not a consequence of many-body interactions, but simply due to spectral overlap between closely spaced excitation pathways. Note that, we considered a reduced contribution from the ESA associated with the feature \textbf{B} for the simulations shown in Fig.~\ref{fig:simul} (b) and (d). While the simulation reproduces the experimental trends very well, we highlight that the model assumes a constant and energy-independent coupling with the continuum and no inhomogeneous broadening, both which have to be rigorously considered for a more true reproduction.

%Notably, the spectral intensity of \textbf{C} reduces at lower coupling strengths, see Fig. S4 in the SI for simulations with lower $q$ values. When the exciton and free carrier states move closer  We show the simulation for distinct coupling conditions associated with $V$ of 20\,meV and 40\,mev.  and further details are in the supplementary information.
%Notably, the model considers a constant distribution for the continuum energy-independent coupling with continuum and dephasing rates. Despite these assumptions, this simple model qualitatively reproduces the main features of the experimentally observed spectra. Additionally, we ignored the inhomogeneous nature of the exciton, which we believe explains the discrepancy when compared with the experimental \textbf{A} feature which elongates towards low energy and could be interpreted as inhomogeneous broadening. 
%The effect of temperature is accounted for by decreasing the difference in energy from 40\,meV to 20\,meV (Fig~\ref{fig:simul} (b) and (d)), and in order to match qualitatively the experimental data we artificially decrease the contribution of \textbf{B}.
%Again, we observe a clear tail that corresponds to the feature \textbf{C}.
%Additionally, the dispersive-like lineshape is reproduced by the overlap of the strong ESA signature with the diagonal features.

We now discuss the evolution of the rephasing spectra, measured at 10\,K with population time (waiting time between pulse 2 and 3), shown in Fig~\ref{fig:fig_pop}. From the absolute maps, Figs~\ref{fig:fig_pop}(a-e), we observe a clear red-shift in the peak of the diagonal feature, by approximately 20\,meV in 90\,fs. %in the features along the diagonal a decrease from the carrier states in comparison with the exciton state. 
The evolution of the spectral intensity along the diagonal can be interpreted as a signature of the transfer of population from the higher energy carrier state to the lower energy exciton state. This transfer is also accompanied by the appearance of a positive cross-peak labeled \textbf{D} in Fig.~\ref{fig:fig_pop}. The photo-generated population after the first two excitation pulses, $\rho_{22} = \ket{k_2}\bra{k_2}$ relaxes to $\rho_{11} = \ket{k_1}\bra{k_1}$, which signifies the population of the lower excitonic state.
The third pulse then generates the coherences of the form $\ket{k_1}\bra{0}$ or $\ket{(k_1k_2)}\bra{k_2}$. These two terms correspond to features \textbf{D} and \textbf{B}, which are accordingly enhanced. In Fig~\ref{fig:fig_pop}(f), we observe that \textbf{A} has a higher intensity than \textbf{B} which supports our assignment of \textbf{A} and \textbf{B} to coupled state since the population evolution creates an asymmetry in the population on the exciton and carrier edge states. Importantly, the observed population relaxation time of 90\,fs is notably faster than the thermalization timescales reported for strongly confined colloidal nanocrystals~\cite{Yu2021}, but comparable to the charge carrier relaxation in bulk \ce{CsPbBr3} crystals~\cite{Nguyen2019, Schlipf2018}. While a longer thermalization process may still be present, the existence of the sub-100fs component may indicate an alleviation of the phonon-bottleneck effects that are ubiquitous to the strongly confined electronic systems. Importantly, there is a dominance of many-body scattering processes in the probed time ranges that drive the relaxation process. 
%We expect the electronic confinement in the isolated PNCs to result in phonon bottleneck effects,  which slow down the carrier thermalization. % recombination and we expect them to be dependent on the level of confinement. 
%The nanocrystal assembly probed here however exhibits ultrafast thermalization indicating the absence of such bottleneck effects and importantly strong presence of multiple scattering events. 
We presume that such many-body scattering events are plausible due to the strong interconnectivity and thus electronic coupling between the PNCs within the assembly. This indicates that the photophysical dynamics are somewhat similar to bulk semiconductors where multiphoton scattering induces effective carrier thermalization~\cite{Schlipf2018}. While a similar coherent nonlinear response was measured by Jha~et~al~\cite{Jha2018} also in the case of bulk perovskites, including signatures for exciton-carrier coupled state, a very prominent difference can be noted. In the bulk \ce{MAPbI3}, the low binding energy exciton was seen to be dissociated in ultrafast timescales to populate the free carrier states, experimentally perceived as a reduction in the inhomogeneous broadening~\cite{Jha2018}. Instead, here we observe an effective population of the lower energy emissive excitonic state in these nanocrystal assemblies. 

\section{\label{sec:Discussion}Conclusion}

In summary, we have presented a study of the exciton-carrier coupling dynamics in a assembly of \ce{FAPbBr3} nanocrystals in a silica scaffold through 2D coherent spectroscopy. The experimentally determined optical signatures indicate coherent coupling of the excitonic and free carrier states in these material systems. This is particularly evidenced by the characteristic Fano-like lineshape in the two-dimensional rephasing spectra.  % that excitons and free carriers are coherently in this type of materials system. 
In addition, we find evidence for the excited state features associated with a coupled exciton-carrier state, similar to what was suggested in bulk metal halide perovskites~\cite{Jha2018}. %In addition, we observe characteristic absorption features (\textbf{A}, \textbf{B}) assigned in this work to coupled exciton-carrier states are similar to those reported for the bulk material \ce{MAPbI3}~\cite{Jha2018}. 
We also observe an effective population of the excitonic state following an ultrafast thermalization process in sub-100\,fs timescale. Our results provide a comprehensive description of the ultrafast excitation dynamics that are driven by coherent interactions between photo-generated carriers and excitonic states in a solid-state assembly of perovskite nanocrystals.    
%However, different from our observation as the population evolves with time, they observe an ultrafast dissociation of excitons as a decrease in the inhomogeneous linewidth.
%Instead, in the low-temperature measurements, where the exciton state is differentiable, we observe a transfer towards the exciton as has been observed for nanocrystals in the past~\cite{Yu2021}, manifested by the increase of the cross-peak \textbf{D}. In addition, we observe signatures of Fano-like interference manifested as a vertical stripe along the $\hbar \omega_1$ axis, signature \textbf{C}.
%Our results describe the dynamics between optical excitations and charges in an intermediate confined \ce{FAPbBr3} nanocrystal assembly, this is of utmost importance to incorporate this class of materials into optoelectronic applications. \\

\begin{acknowledgments}
A.R.S.K. acknowledges the start-up funds provided by Wake Forest University and funding from the Center for Functional Materials and the Office of Research and Sponsored Programs at WFU. The authors thank Professor Carlos Silva for giving access to the optical instrumentation and for insightful discussions. The optical instrumentation was supported by the National Science Foundation (DMR-1904293). The experimental data collection, analysis, and the writing of corresponding manuscript sections by E.R.G. were supported by the National Science Foundation (DMR-2019444). 
D.O.T. acknowledges financial support from the European Union’s Horizon 2020 research and innovation program under the Marie Skłodowska-Curie grant agreement No 956270.
H.M. is thankful for the financial support received from the Spanish Ministry of Science and Innovation-Agencia Estatal de Investigación (MICIN-AEI) under grants PID2020-116593RB-I00, funded by MCIN/AEI/ 10.13039/501100011033, from the Junta de Andalucía under grant P18-RT-2291 (FEDER/UE) and from the Innovative Training Network Persephone ITN, funded by the European Union’s Horizon 2020 research.
and innovation program under the Marie Skłodowska-Curie grant agreement No 956270. 

\end{acknowledgments}

\section*{Author Contributions}

The measurements were performed by E.R-G., D.O.T., and K.A.K. under the supervision of A.R.S.K. The samples were prepared by CRP under the supervision of J.F.G-L., M.E.C., and H.M. E.R.G. wrote the original draft and all the authors contributed to the editing of the manuscript. 

%Appendixes here

%\bibliography{BIB}% Produces the bibliography via BibTeX.
\providecommand{\noopsort}[1]{}\providecommand{\singleletter}[1]{#1}%

\clearpage
\newpage

\onecolumngrid
\textbf{\Large{Supplementary Information: Exciton-carrier coupling in a metal halide perovskite nanocrystal assembly probed by two-dimensional coherent spectroscopy}}
\vspace{1cm}%
\twocolumngrid

\section{Sample preparation and characterization}

\subsection{Sample preparation}
\paragraph{Materials} For precursor sample preparation formamidinium bromide (\ce{FABr}, GreatCell Solar Materials, 99.9\%), Lead (II) bromide (\ce{PbBr2}, TCI, 99.99\%), dimethylsulfoxide (DMSO, Merck, anhydrous 99.8\%), methanol (\ce{MeOH}, VWR, 98\%) and chlorobenzene (\ce{CB}, Merck, 99.9\%) were employed without additional purification steps. \\

\paragraph{Preparation of \ce{SiO2} nanoparticles porous scaffold} A colloidal suspension of 30\,nm \ce{SiO2} nanoparticles (34\% w/v in H2O, LUDOX-TMA, Sigma-Aldrich) was obtained by diluting with methanol to 3\% w/v. Low-fluorescence glass substrates were dip coated with the diluted suspension with a withdrawal speed of 120\,mm/min. The deposition was repeated 15 times to achieve a thickness of roughly 1\,µm. The porous matrix was consequently annealed at 450ºC for 30\,min to remove any residual organic component and improve its mechanical stability. \\ 

\paragraph{Synthesis of \ce{FAPbBr3} nanocrystals within nanoporous silica scaffold} \ce{FABr} and \ce{PbBr2} powders were dissolved in \ce{DMSO} in a 1:1 molar ratio with 30\% concentration to obtain a perovskite solution precursor. The solution was infiltrated into the voids of the porous matrix by spin-coating with 5000\,rpm for 60\,s. The sample was anneahled at 100ºC for 1\,hour to crystalize \ce{FAPbBr3} nanocrystals within the pores of the matrix. All preparation steps of the synthesis of \ce{FAPbBr3} nanocrystals were performed in a protected \ce{N2} environment of a glovebox. 

\subsection{Sample characterization}
The absolute PL quantum yield (PLQY) was measured using a commercial flourometer (Edinburgh FLS1000) and an integrating sphere accessory employing a continuous light excitation source with $\lambda_{exc} = 450 \text{\,nm}$. The cw-PLQY was determined to be  1.3\,\%.
Perovskite pore filling fraction was determined to be 14.5\,\% of the pore volume by Atomic Emission Spectroscopy by Inductively Coupled Plasma Atomic Emission Spectroscopy (ICP-AES), using 10 mL water solutions of soaked FAPbBr3 QD-SiO2 films for 1 hour (complete extraction of perovskite in the film) in an iCAP 7200 ICP-AES Duo (ThermoFisher Scientific) equipment.

\begin{figure}
    \centering
    \includegraphics[width=0.5\linewidth]{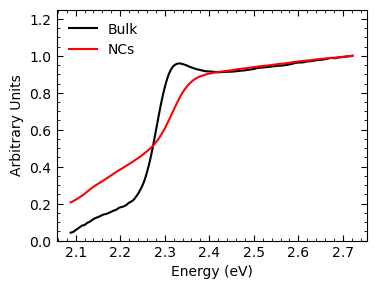}
    \caption{Comparison between the absorption spectra of \ce{FAPbBr3} bulk semiconductor and \ce{FAPbBr3} nanocrystal (NCs) assembly thin films measured at room temperature.}
    \label{fig:Abs}
\end{figure}

\section{2D coherent electronic spectroscopy.}

\begin{figure}
    \centering
    \includegraphics[width=0.7\linewidth]{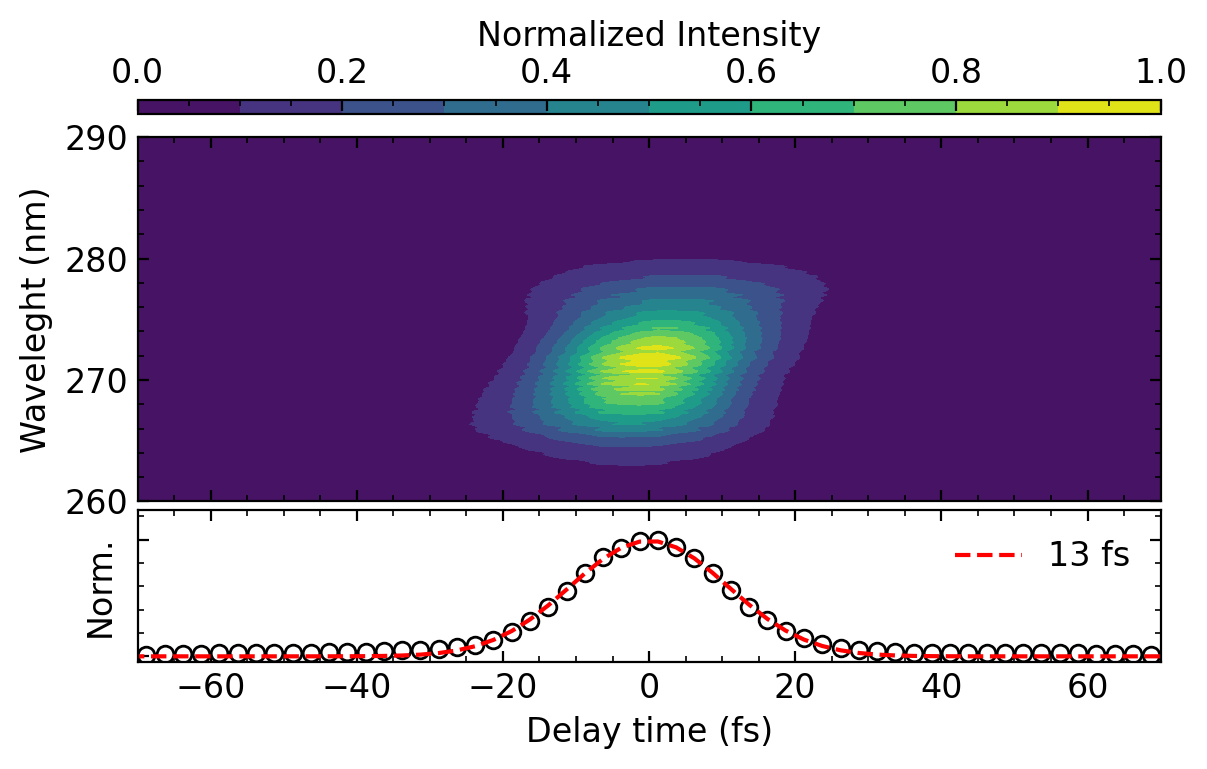}
    \caption{X-Frog characterization of the pulse duration fitted to a Gaussian function.}
    \label{fig:frog}
\end{figure}

We used the same system as previously implemented~\cite{Thouin2018, Thouin2019PRR}. The four beams share a common path for passive phase stabilization and with an SLM we apply the time delay and compress the pulses. It was developed by Turner and coworkers~\cite{Turner2011}. The pulses were individually compressed using a home-built implementation of a chirp scan~\cite{Loriot2013}. 
%The geometry of the experiment is summarized in Fig.~\ref{fig:ExpScheme}. 
Our home-built third-harmonic-pumped non-collinear optical parametric amplifier is pumped by a portion of the output pf a commercial ultrafast laser system (Light Conversion Pharos) at a 100-kHz repetition rate. The pulse has wavelength of 1030\,nm and a duration of 220\,fs. The resulting pulse duration, after the NOPA and the compression, was below 15\,fs full-width at half-maximum, as measured by second-harmonic generation cross-frequency-resolved optical gating (SHG-XFROG), shown in Fig.~\ref{fig:frog}.
All measurements were carried out in a vibration-free closed-cycle cryostat (Montana Instruments).

\onecolumngrid

\subsection{Data Analysis}
\subsubsection{Spectra interferometry}

The contents of this section have already been discussed in previous literature~\cite{Turner2011}. For the sake of reproducibility and completeness, we discuss the data analysis and the specifics of our implementation. We start from a general form of the emitted electric field when the pulses have an arbitrary phase has a general form:

\begin{equation}
    E_{sig} = A(\omega)[e^{-i\phi_{sig}(\omega)-i\Delta\Phi}+e^{+i\phi_{sig}(\omega)+i\Delta\Phi}].
\end{equation}
The global phase $\Delta \Phi$ is defined as:
\begin{equation}
    \Delta \Phi = \delta\phi_{LO}-\delta\phi_{a}+\delta\phi_{b}+\delta\phi_{c}.
\end{equation}

\noindent
The detected heterodyne signal is:
\begin{widetext}

\begin{equation}
    I(\omega) = |E_{sig}+E_{LO}|^2.
\end{equation}

\begin{equation}
    I(\omega) = |E_{sig}|^2+|E_{LO}|^2+ E^*_{sig}E_{LO}+E_{sig}E^*_{LO}.
\end{equation}

\begin{equation}
    I_{XT}(\omega) = 2 A(\omega)A_{LO}(\omega)\left[e^{-i[\phi_{sig}(\omega)+\Delta\Phi-\phi_{LO}(\omega)]}+e^{+i[\phi_{sig}(\omega)+\Delta\Phi-\phi_{LO}(\omega)]}\right].
\end{equation}

\begin{equation}
    I_{XT}(\omega) =  4A(\omega)A_{LO}(\omega)\cos(\phi_{sig}(\omega)+\Delta\Phi-\phi_{LO}(\omega)).
\end{equation}
\end{widetext}

\onecolumngrid
\subsubsection{Post-processing and phasing}
\begin{figure}
    \centering
    \includegraphics[width=0.5\linewidth]{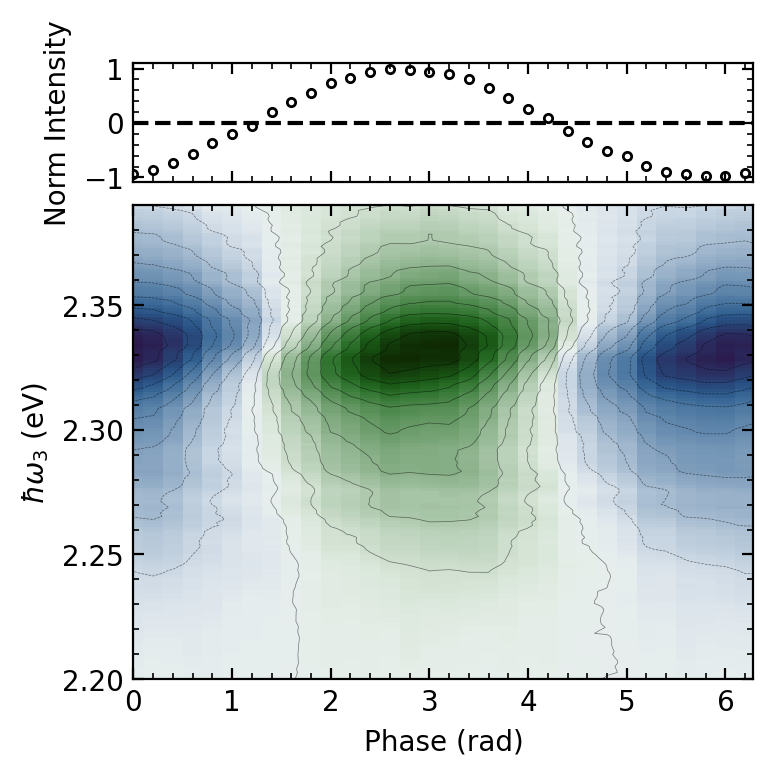}
    \caption{(Top) Cut at 2.32\,eV from the phase scan performed. (Bottom) Complete phase scan obtained for the measurement at 10\,K with a population time of 20\,fs}
    \label{fig:phase}
\end{figure}

\noindent
After the acquisition of the data, we take the inverse Fourier transform (define as $\int d\omega f(\omega)e^{i\omega t}$), remove the negative times numerically, and do a Fourier transform back ($\int dt f(t)e^{-i\omega t}$):

\begin{equation}
    FT^{-1}(I_{XT}(\omega)) = B^+(t)+B^-(t).
\end{equation}

\begin{equation}
    FT(B^+(t)) = A(\omega)A_{LO}(\omega)e^{-i[\phi_{sig}(\omega)+\Delta\Phi-\phi_{LO}(\omega)]}.
\end{equation}

\noindent
To isolate the emitted electric field we need to know $\phi_{LO}(\omega)$ such that it can be removed, we assume the $\phi_{LO}(\omega) = (\omega - \omega_{LO})\tau_{LO}$:

\begin{equation}
    E_{sig}(\omega) = \frac{A(\omega)A_{LO}(\omega)e^{-i[\phi_{sig}(\omega)+\Delta\Phi-\phi_{LO}(\omega)]}}{A_{LO}(\omega)e^{-\phi_{LO}(\omega)}} = E_{sig}(\omega) = A(\omega)e^{-i\phi_{sig}(\omega)}e^{-i\Delta \Phi}
\end{equation}

\noindent
To correct the phase of the measured four-wave-mixing signal through heterodyne detection. We scan the phase of one of the pulses (Beam 4 in this case) using the spatial light modulator, then:

\begin{equation}
    E_{sig}(\omega) = A(\omega)e^{-i\phi_{sig}(\omega)}e^{-i\Delta \Phi+i\phi_{scanned}}
\end{equation}

\begin{equation}
    \Re(E_{sig}(\omega)) = A(\omega)\cos(\Delta \Phi - \phi_{scanned})
\end{equation}

The obtained map after scanning is shown in Fig.~\ref{fig:phase}.

\begin{figure}
    \centering
    \includegraphics{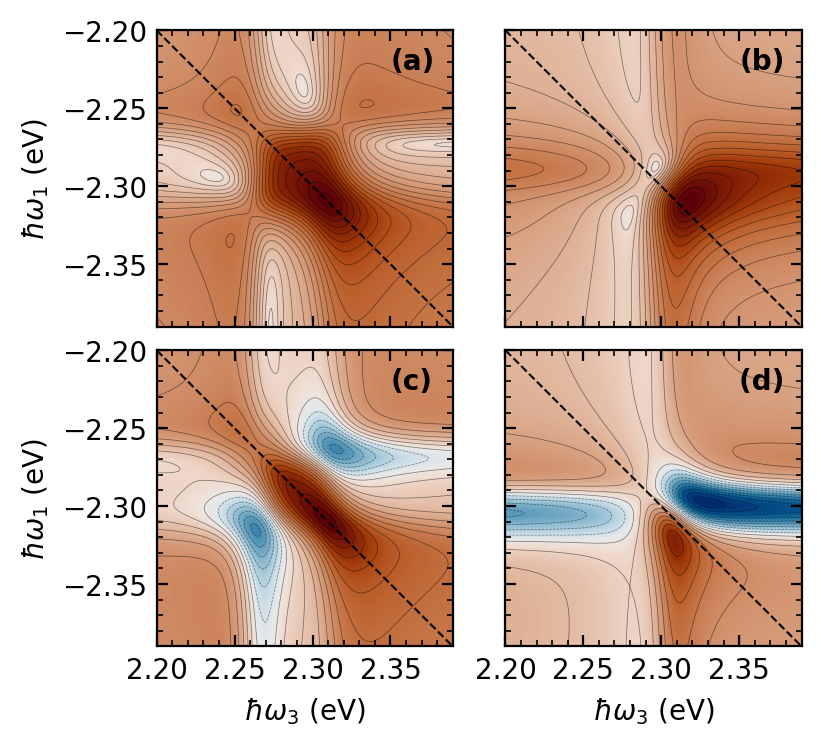}
    \caption{Simulated rephasing two-dimensional maps for two states separated by (a) 40\,meV and (b) 20\,meV, which include the contributions from coupled excited state pathways. The coupling with the continuum was set to $q=0.5$. Further simulation details can be found in the supporting information.}
    \label{fig:SimulationCondition2}
\end{figure}

\subsection{Fano-like interference expresions}

The expressions were derived by Filkenstein \textit{et al} previously~\cite{Fano1, Fano2}. Here, we just show the expressions used in the main text. The rephasing ground state bleach, stimulated emission expressions, and excited state absorption are given by:
\begin{equation}
    R_{GB}(\epsilon_t, \epsilon_{\tau}) = (n\,\Gamma \pi (q^2+1))^2 \frac{e^{-(2\gamma_e+\eta)T}}{(\epsilon_{\tau}-i)(\epsilon_t+i)}
\end{equation}

\begin{equation}
    R_{SE}(\epsilon_t, \epsilon_{\tau}) = (n \pi)^2 \left(\frac{\Gamma (q-i)^2}{\epsilon_t+i}-i\right)\left(\frac{\Gamma (q+i)^2}{\epsilon_{\tau}-i}+i\right)
\end{equation}

\begin{equation}
    R_{ESA}(\epsilon_t, \epsilon_{\tau}) = -(n\,\Gamma \pi (q^2+1))^2 \frac{e^{-(2\gamma_e+\eta)T}}{(\epsilon_{\tau}-i)(\epsilon_t+i)}-(n \pi)^2 \left(\frac{\Gamma (q-i)^2}{\epsilon_t+i}-1\right)\left(\frac{\Gamma (q+i)^2}{\epsilon_{\tau}-i}+i\right)
\end{equation} 

\noindent
In the expressions above, $\epsilon_{\tau} = (\omega_1-\omega_e)/(\gamma_c + \gamma_e)$ and $\epsilon_{t} = (\omega_3-\omega_{e'})/(\gamma_c + \gamma_{e'})$  where $\hbar\omega_e$ corresponds to the frequency of the transition and $\hbar\gamma_e$ is the dephasing. Note that for the ESA cross peaks $\hbar\omega_e \neq \hbar\omega_{e'}$. Similarly, $\hbar\gamma_c$ is the continuum dephasing. For all simulations, we chose $\ hbar\gamma_e = \hbar\gamma_c = 0.01$\,eV. We also chose a waiting time ($t_2$ in the main text) $T = 0$ as did not include any kinetic model to describe their population dynamics. As we chose $T=0$ the choice of $\eta$ is not relevant. As stated in the main text, the coupling strength is determined by the parameter $q$, $q = \frac{\mu_e}{\mu_c \pi V}$ where $V$ is the coupling constant between the continuum of states and the discrete transition. $\mu_e$ and $\mu_c$ correspond to the transition moments of the discrete and continuum transition respectively. 

%\printbibliography
%\bibliography{BibSI}

\begin{thebibliography}{53}%
\makeatletter
\providecommand \@ifxundefined [1]{%
 \@ifx{#1\undefined}
}%
\providecommand \@ifnum [1]{%
 \ifnum #1\expandafter \@firstoftwo
 \else \expandafter \@secondoftwo
 \fi
}%
\providecommand \@ifx [1]{%
 \ifx #1\expandafter \@firstoftwo
 \else \expandafter \@secondoftwo
 \fi
}%
\providecommand \natexlab [1]{#1}%
\providecommand \enquote  [1]{``#1''}%
\providecommand \bibnamefont  [1]{#1}%
\providecommand \bibfnamefont [1]{#1}%
\providecommand \citenamefont [1]{#1}%
\providecommand \href@noop [0]{\@secondoftwo}%
\providecommand \href [0]{\begingroup \@sanitize@url \@href}%
\providecommand \@href[1]{\@@startlink{#1}\@@href}%
\providecommand \@@href[1]{\endgroup#1\@@endlink}%
\providecommand \@sanitize@url [0]{\catcode `\\12\catcode `\$12\catcode
  `\&12\catcode `\#12\catcode `\^12\catcode `\_12\catcode `\%12\relax}%
\providecommand \@@startlink[1]{}%
\providecommand \@@endlink[0]{}%
\providecommand \url  [0]{\begingroup\@sanitize@url \@url }%
\providecommand \@url [1]{\endgroup\@href {#1}{\urlprefix }}%
\providecommand \urlprefix  [0]{URL }%
\providecommand \Eprint [0]{\href }%
\providecommand \doibase [0]{http://dx.doi.org/}%
\providecommand \selectlanguage [0]{\@gobble}%
\providecommand \bibinfo  [0]{\@secondoftwo}%
\providecommand \bibfield  [0]{\@secondoftwo}%
\providecommand \translation [1]{[#1]}%
\providecommand \BibitemOpen [0]{}%
\providecommand \bibitemStop [0]{}%
\providecommand \bibitemNoStop [0]{.\EOS\space}%
\providecommand \EOS [0]{\spacefactor3000\relax}%
\providecommand \BibitemShut  [1]{\csname bibitem#1\endcsname}%
\let\auto@bib@innerbib\@empty
%</preamble>
\bibitem [{\citenamefont {Swarnkar}\ \emph {et~al.}(2016)\citenamefont
  {Swarnkar}, \citenamefont {Marshall}, \citenamefont {Sanehira}, \citenamefont
  {Chernomordik}, \citenamefont {Moore}, \citenamefont {Christians},
  \citenamefont {Chakrabarti},\ and\ \citenamefont {Luther}}]{Abhishek2016}%
  \BibitemOpen
  \bibfield  {author} {\bibinfo {author} {\bibfnamefont {A.}~\bibnamefont
  {Swarnkar}}, \bibinfo {author} {\bibfnamefont {A.~R.}\ \bibnamefont
  {Marshall}}, \bibinfo {author} {\bibfnamefont {E.~M.}\ \bibnamefont
  {Sanehira}}, \bibinfo {author} {\bibfnamefont {B.~D.}\ \bibnamefont
  {Chernomordik}}, \bibinfo {author} {\bibfnamefont {D.~T.}\ \bibnamefont
  {Moore}}, \bibinfo {author} {\bibfnamefont {J.~A.}\ \bibnamefont
  {Christians}}, \bibinfo {author} {\bibfnamefont {T.}~\bibnamefont
  {Chakrabarti}}, \ and\ \bibinfo {author} {\bibfnamefont {J.~M.}\ \bibnamefont
  {Luther}},\ }\href {\doibase 10.1126/science.aag2700} {\bibfield  {journal}
  {\bibinfo  {journal} {Science}\ }\textbf {\bibinfo {volume} {354}},\ \bibinfo
  {pages} {92} (\bibinfo {year} {2016})}\BibitemShut {NoStop}%
\bibitem [{\citenamefont {Chiba}\ \emph {et~al.}(2018)\citenamefont {Chiba},
  \citenamefont {Hayashi}, \citenamefont {Ebe}, \citenamefont {Hoshi},
  \citenamefont {Sato}, \citenamefont {Sato}, \citenamefont {Pu}, \citenamefont
  {Ohisa},\ and\ \citenamefont {Kido}}]{Chiba2018}%
  \BibitemOpen
  \bibfield  {author} {\bibinfo {author} {\bibfnamefont {T.}~\bibnamefont
  {Chiba}}, \bibinfo {author} {\bibfnamefont {Y.}~\bibnamefont {Hayashi}},
  \bibinfo {author} {\bibfnamefont {H.}~\bibnamefont {Ebe}}, \bibinfo {author}
  {\bibfnamefont {K.}~\bibnamefont {Hoshi}}, \bibinfo {author} {\bibfnamefont
  {J.}~\bibnamefont {Sato}}, \bibinfo {author} {\bibfnamefont {S.}~\bibnamefont
  {Sato}}, \bibinfo {author} {\bibfnamefont {Y.-J.}\ \bibnamefont {Pu}},
  \bibinfo {author} {\bibfnamefont {S.}~\bibnamefont {Ohisa}}, \ and\ \bibinfo
  {author} {\bibfnamefont {J.}~\bibnamefont {Kido}},\ }\href {\doibase
  10.1038/s41566-018-0260-y} {\bibfield  {journal} {\bibinfo  {journal} {Nat.
  Photonics}\ }\textbf {\bibinfo {volume} {12}},\ \bibinfo {pages} {681}
  (\bibinfo {year} {2018})}\BibitemShut {NoStop}%
\bibitem [{\citenamefont {Dirin}\ \emph {et~al.}(2016)\citenamefont {Dirin},
  \citenamefont {Protesescu}, \citenamefont {Trummer}, \citenamefont
  {Kochetygov}, \citenamefont {Yakunin}, \citenamefont {Krumeich},
  \citenamefont {Stadie},\ and\ \citenamefont
  {Kovalenko}}]{Dirin2016Harnessing}%
  \BibitemOpen
  \bibfield  {author} {\bibinfo {author} {\bibfnamefont {D.~N.}\ \bibnamefont
  {Dirin}}, \bibinfo {author} {\bibfnamefont {L.}~\bibnamefont {Protesescu}},
  \bibinfo {author} {\bibfnamefont {D.}~\bibnamefont {Trummer}}, \bibinfo
  {author} {\bibfnamefont {I.~V.}\ \bibnamefont {Kochetygov}}, \bibinfo
  {author} {\bibfnamefont {S.}~\bibnamefont {Yakunin}}, \bibinfo {author}
  {\bibfnamefont {F.}~\bibnamefont {Krumeich}}, \bibinfo {author}
  {\bibfnamefont {N.~P.}\ \bibnamefont {Stadie}}, \ and\ \bibinfo {author}
  {\bibfnamefont {M.~V.}\ \bibnamefont {Kovalenko}},\ }\href {\doibase
  10.1021/acs.nanolett.6b02688} {\bibfield  {journal} {\bibinfo  {journal}
  {Nano Lett.}\ }\textbf {\bibinfo {volume} {16}},\ \bibinfo {pages} {5866}
  (\bibinfo {year} {2016})}\BibitemShut {NoStop}%
\bibitem [{\citenamefont {Malgras}\ \emph {et~al.}(2016)\citenamefont
  {Malgras}, \citenamefont {Tominaka}, \citenamefont {Ryan}, \citenamefont
  {Henzie}, \citenamefont {Takei}, \citenamefont {Ohara},\ and\ \citenamefont
  {Yamauchi}}]{Malgras2016Observation}%
  \BibitemOpen
  \bibfield  {author} {\bibinfo {author} {\bibfnamefont {V.}~\bibnamefont
  {Malgras}}, \bibinfo {author} {\bibfnamefont {S.}~\bibnamefont {Tominaka}},
  \bibinfo {author} {\bibfnamefont {J.~W.}\ \bibnamefont {Ryan}}, \bibinfo
  {author} {\bibfnamefont {J.}~\bibnamefont {Henzie}}, \bibinfo {author}
  {\bibfnamefont {T.}~\bibnamefont {Takei}}, \bibinfo {author} {\bibfnamefont
  {K.}~\bibnamefont {Ohara}}, \ and\ \bibinfo {author} {\bibfnamefont
  {Y.}~\bibnamefont {Yamauchi}},\ }\href {\doibase 10.1021/jacs.6b05608}
  {\bibfield  {journal} {\bibinfo  {journal} {J. Am. Chem. Soc.}\ }\textbf
  {\bibinfo {volume} {138}},\ \bibinfo {pages} {13874} (\bibinfo {year}
  {2016})}\BibitemShut {NoStop}%
\bibitem [{\citenamefont {Rubino}\ \emph {et~al.}(2020)\citenamefont {Rubino},
  \citenamefont {Cali{\`o}}, \citenamefont {Garc{\'\i}a-Bennett}, \citenamefont
  {Calvo},\ and\ \citenamefont {M{\'\i}guez}}]{rubino2020mesoporous}%
  \BibitemOpen
  \bibfield  {author} {\bibinfo {author} {\bibfnamefont {A.}~\bibnamefont
  {Rubino}}, \bibinfo {author} {\bibfnamefont {L.}~\bibnamefont {Cali{\`o}}},
  \bibinfo {author} {\bibfnamefont {A.}~\bibnamefont {Garc{\'\i}a-Bennett}},
  \bibinfo {author} {\bibfnamefont {M.~E.}\ \bibnamefont {Calvo}}, \ and\
  \bibinfo {author} {\bibfnamefont {H.}~\bibnamefont {M{\'\i}guez}},\
  }\href@noop {} {\bibfield  {journal} {\bibinfo  {journal} {Adv. Opt. Mater.}\
  }\textbf {\bibinfo {volume} {8}},\ \bibinfo {pages} {1901868} (\bibinfo
  {year} {2020})}\BibitemShut {NoStop}%
\bibitem [{\citenamefont {Srimath~Kandada}\ \emph {et~al.}(2016)\citenamefont
  {Srimath~Kandada}, \citenamefont {Neutzner}, \citenamefont {D’Innocenzo},
  \citenamefont {Tassone}, \citenamefont {Gandini}, \citenamefont {Akkerman},
  \citenamefont {Prato}, \citenamefont {Manna}, \citenamefont {Petrozza},\ and\
  \citenamefont {Lanzani}}]{kandada2016}%
  \BibitemOpen
  \bibfield  {author} {\bibinfo {author} {\bibfnamefont {A.~R.}\ \bibnamefont
  {Srimath~Kandada}}, \bibinfo {author} {\bibfnamefont {S.}~\bibnamefont
  {Neutzner}}, \bibinfo {author} {\bibfnamefont {V.}~\bibnamefont
  {D’Innocenzo}}, \bibinfo {author} {\bibfnamefont {F.}~\bibnamefont
  {Tassone}}, \bibinfo {author} {\bibfnamefont {M.}~\bibnamefont {Gandini}},
  \bibinfo {author} {\bibfnamefont {Q.~A.}\ \bibnamefont {Akkerman}}, \bibinfo
  {author} {\bibfnamefont {M.}~\bibnamefont {Prato}}, \bibinfo {author}
  {\bibfnamefont {L.}~\bibnamefont {Manna}}, \bibinfo {author} {\bibfnamefont
  {A.}~\bibnamefont {Petrozza}}, \ and\ \bibinfo {author} {\bibfnamefont
  {G.}~\bibnamefont {Lanzani}},\ }\href {\doibase 10.1021/jacs.6b06463}
  {\bibfield  {journal} {\bibinfo  {journal} {J. Am. Chem. Soc.}\ }\textbf
  {\bibinfo {volume} {138}},\ \bibinfo {pages} {13604} (\bibinfo {year}
  {2016})}\BibitemShut {NoStop}%
\bibitem [{\citenamefont {Jariwala}\ \emph {et~al.}(2021)\citenamefont
  {Jariwala}, \citenamefont {Burke}, \citenamefont {Dunfield}, \citenamefont
  {Shallcross}, \citenamefont {Taddei}, \citenamefont {Wang}, \citenamefont
  {Eperon}, \citenamefont {Armstrong}, \citenamefont {Berry},\ and\
  \citenamefont {Ginger}}]{Jariwala2021}%
  \BibitemOpen
  \bibfield  {author} {\bibinfo {author} {\bibfnamefont {S.}~\bibnamefont
  {Jariwala}}, \bibinfo {author} {\bibfnamefont {S.}~\bibnamefont {Burke}},
  \bibinfo {author} {\bibfnamefont {S.}~\bibnamefont {Dunfield}}, \bibinfo
  {author} {\bibfnamefont {R.~C.}\ \bibnamefont {Shallcross}}, \bibinfo
  {author} {\bibfnamefont {M.}~\bibnamefont {Taddei}}, \bibinfo {author}
  {\bibfnamefont {J.}~\bibnamefont {Wang}}, \bibinfo {author} {\bibfnamefont
  {G.~E.}\ \bibnamefont {Eperon}}, \bibinfo {author} {\bibfnamefont {N.~R.}\
  \bibnamefont {Armstrong}}, \bibinfo {author} {\bibfnamefont {J.~J.}\
  \bibnamefont {Berry}}, \ and\ \bibinfo {author} {\bibfnamefont {D.~S.}\
  \bibnamefont {Ginger}},\ }\href {\doibase 10.1021/acs.chemmater.1c00848}
  {\bibfield  {journal} {\bibinfo  {journal} {Chem. Mat.}\ }\textbf {\bibinfo
  {volume} {33}},\ \bibinfo {pages} {5035} (\bibinfo {year}
  {2021})}\BibitemShut {NoStop}%
\bibitem [{\citenamefont {Rubino}\ \emph {et~al.}(2021)\citenamefont {Rubino},
  \citenamefont {Caliò}, \citenamefont {Calvo},\ and\ \citenamefont
  {Míguez}}]{Rubino2021}%
  \BibitemOpen
  \bibfield  {author} {\bibinfo {author} {\bibfnamefont {A.}~\bibnamefont
  {Rubino}}, \bibinfo {author} {\bibfnamefont {L.}~\bibnamefont {Caliò}},
  \bibinfo {author} {\bibfnamefont {M.~E.}\ \bibnamefont {Calvo}}, \ and\
  \bibinfo {author} {\bibfnamefont {H.}~\bibnamefont {Míguez}},\ }\href
  {\doibase https://doi.org/10.1002/solr.202100204} {\bibfield  {journal}
  {\bibinfo  {journal} {Sol. RRL}\ }\textbf {\bibinfo {volume} {5}},\ \bibinfo
  {pages} {2100204} (\bibinfo {year} {2021})}\BibitemShut {NoStop}%
\bibitem [{\citenamefont {Romero-P{\'e}rez}\ \emph {et~al.}(2022)\citenamefont
  {Romero-P{\'e}rez}, \citenamefont {Rubino}, \citenamefont {Cali{\`o}},
  \citenamefont {Calvo},\ and\ \citenamefont
  {M{\'\i}guez}}]{romero2022optoelectronic}%
  \BibitemOpen
  \bibfield  {author} {\bibinfo {author} {\bibfnamefont {C.}~\bibnamefont
  {Romero-P{\'e}rez}}, \bibinfo {author} {\bibfnamefont {A.}~\bibnamefont
  {Rubino}}, \bibinfo {author} {\bibfnamefont {L.}~\bibnamefont {Cali{\`o}}},
  \bibinfo {author} {\bibfnamefont {M.~E.}\ \bibnamefont {Calvo}}, \ and\
  \bibinfo {author} {\bibfnamefont {H.}~\bibnamefont {M{\'\i}guez}},\
  }\href@noop {} {\bibfield  {journal} {\bibinfo  {journal} {Adv. Opt. Mater.}\
  }\textbf {\bibinfo {volume} {10}},\ \bibinfo {pages} {2102112} (\bibinfo
  {year} {2022})}\BibitemShut {NoStop}%
\bibitem [{\citenamefont {Wehner}\ \emph {et~al.}(1996)\citenamefont {Wehner},
  \citenamefont {Steinbach},\ and\ \citenamefont {Wegener}}]{Wehner1996}%
  \BibitemOpen
  \bibfield  {author} {\bibinfo {author} {\bibfnamefont {M.~U.}\ \bibnamefont
  {Wehner}}, \bibinfo {author} {\bibfnamefont {D.}~\bibnamefont {Steinbach}}, \
  and\ \bibinfo {author} {\bibfnamefont {M.}~\bibnamefont {Wegener}},\ }\href
  {\doibase 10.1103/PhysRevB.54.R5211} {\bibfield  {journal} {\bibinfo
  {journal} {Phys. Rev. B}\ }\textbf {\bibinfo {volume} {54}},\ \bibinfo
  {pages} {R5211} (\bibinfo {year} {1996})}\BibitemShut {NoStop}%
\bibitem [{\citenamefont {Webber}\ \emph {et~al.}(2016)\citenamefont {Webber},
  \citenamefont {Wilmer}, \citenamefont {Liu}, \citenamefont {Dobrowolska},
  \citenamefont {Furdyna}, \citenamefont {Bristow},\ and\ \citenamefont
  {Hall}}]{Webber2016}%
  \BibitemOpen
  \bibfield  {author} {\bibinfo {author} {\bibfnamefont {D.}~\bibnamefont
  {Webber}}, \bibinfo {author} {\bibfnamefont {B.~L.}\ \bibnamefont {Wilmer}},
  \bibinfo {author} {\bibfnamefont {X.}~\bibnamefont {Liu}}, \bibinfo {author}
  {\bibfnamefont {M.}~\bibnamefont {Dobrowolska}}, \bibinfo {author}
  {\bibfnamefont {J.~K.}\ \bibnamefont {Furdyna}}, \bibinfo {author}
  {\bibfnamefont {A.~D.}\ \bibnamefont {Bristow}}, \ and\ \bibinfo {author}
  {\bibfnamefont {K.~C.}\ \bibnamefont {Hall}},\ }\href {\doibase
  10.1103/PhysRevB.94.155450} {\bibfield  {journal} {\bibinfo  {journal} {Phys.
  Rev. B}\ }\textbf {\bibinfo {volume} {94}},\ \bibinfo {pages} {155450}
  (\bibinfo {year} {2016})}\BibitemShut {NoStop}%
\bibitem [{\citenamefont {Allan}\ and\ \citenamefont
  {Driel}(1999)}]{Allan1999}%
  \BibitemOpen
  \bibfield  {author} {\bibinfo {author} {\bibfnamefont {G.~R.}\ \bibnamefont
  {Allan}}\ and\ \bibinfo {author} {\bibfnamefont {H.~M.~v.}\ \bibnamefont
  {Driel}},\ }\href {\doibase 10.1103/PhysRevB.59.15740} {\bibfield  {journal}
  {\bibinfo  {journal} {Phys. Rev. B}\ }\textbf {\bibinfo {volume} {59}},\
  \bibinfo {pages} {15740} (\bibinfo {year} {1999})}\BibitemShut {NoStop}%
\bibitem [{\citenamefont {Trinh}\ \emph {et~al.}(2015)\citenamefont {Trinh},
  \citenamefont {Wu}, \citenamefont {Niesner},\ and\ \citenamefont
  {Zhu}}]{Trinh2015}%
  \BibitemOpen
  \bibfield  {author} {\bibinfo {author} {\bibfnamefont {M.~T.}\ \bibnamefont
  {Trinh}}, \bibinfo {author} {\bibfnamefont {X.}~\bibnamefont {Wu}}, \bibinfo
  {author} {\bibfnamefont {D.}~\bibnamefont {Niesner}}, \ and\ \bibinfo
  {author} {\bibfnamefont {X.-Y.}\ \bibnamefont {Zhu}},\ }\href {\doibase
  10.1039/C5TA01093D} {\bibfield  {journal} {\bibinfo  {journal} {J. Mater.
  Chem. A}\ }\textbf {\bibinfo {volume} {3}},\ \bibinfo {pages} {9285}
  (\bibinfo {year} {2015})}\BibitemShut {NoStop}%
\bibitem [{\citenamefont {Jha}\ \emph {et~al.}(2018)\citenamefont {Jha},
  \citenamefont {Duan}, \citenamefont {Tiwari}, \citenamefont {Nayak},
  \citenamefont {Snaith}, \citenamefont {Thorwart},\ and\ \citenamefont
  {Miller}}]{Jha2018}%
  \BibitemOpen
  \bibfield  {author} {\bibinfo {author} {\bibfnamefont {A.}~\bibnamefont
  {Jha}}, \bibinfo {author} {\bibfnamefont {H.-G.}\ \bibnamefont {Duan}},
  \bibinfo {author} {\bibfnamefont {V.}~\bibnamefont {Tiwari}}, \bibinfo
  {author} {\bibfnamefont {P.~K.}\ \bibnamefont {Nayak}}, \bibinfo {author}
  {\bibfnamefont {H.~J.}\ \bibnamefont {Snaith}}, \bibinfo {author}
  {\bibfnamefont {M.}~\bibnamefont {Thorwart}}, \ and\ \bibinfo {author}
  {\bibfnamefont {R.~J.~D.}\ \bibnamefont {Miller}},\ }\href {\doibase
  10.1021/acsphotonics.7b01025} {\bibfield  {journal} {\bibinfo  {journal} {ACS
  Photonics}\ }\textbf {\bibinfo {volume} {5}},\ \bibinfo {pages} {852}
  (\bibinfo {year} {2018})}\BibitemShut {NoStop}%
\bibitem [{\citenamefont {Palmieri}\ \emph {et~al.}(2020)\citenamefont
  {Palmieri}, \citenamefont {Baldini}, \citenamefont {Steinhoff}, \citenamefont
  {Akrap}, \citenamefont {Koll{\'a}r}, \citenamefont {Horv{\'a}th},
  \citenamefont {Forr{\'o}}, \citenamefont {Jahnke},\ and\ \citenamefont
  {Chergui}}]{Palmieri2020Mahan}%
  \BibitemOpen
  \bibfield  {author} {\bibinfo {author} {\bibfnamefont {T.}~\bibnamefont
  {Palmieri}}, \bibinfo {author} {\bibfnamefont {E.}~\bibnamefont {Baldini}},
  \bibinfo {author} {\bibfnamefont {A.}~\bibnamefont {Steinhoff}}, \bibinfo
  {author} {\bibfnamefont {A.}~\bibnamefont {Akrap}}, \bibinfo {author}
  {\bibfnamefont {M.}~\bibnamefont {Koll{\'a}r}}, \bibinfo {author}
  {\bibfnamefont {E.}~\bibnamefont {Horv{\'a}th}}, \bibinfo {author}
  {\bibfnamefont {L.}~\bibnamefont {Forr{\'o}}}, \bibinfo {author}
  {\bibfnamefont {F.}~\bibnamefont {Jahnke}}, \ and\ \bibinfo {author}
  {\bibfnamefont {M.}~\bibnamefont {Chergui}},\ }\href {\doibase
  10.1038/s41467-020-14683-5} {\bibfield  {journal} {\bibinfo  {journal} {Nat.
  Commun.}\ }\textbf {\bibinfo {volume} {11}},\ \bibinfo {pages} {850}
  (\bibinfo {year} {2020})}\BibitemShut {NoStop}%
\bibitem [{\citenamefont {Nguyen}\ \emph {et~al.}(2019)\citenamefont {Nguyen},
  \citenamefont {Timmer}, \citenamefont {Rakita}, \citenamefont {Cahen},
  \citenamefont {Steinhoff}, \citenamefont {Jahnke}, \citenamefont {Lienau},\
  and\ \citenamefont {De~Sio}}]{Nguyen2019}%
  \BibitemOpen
  \bibfield  {author} {\bibinfo {author} {\bibfnamefont {X.~T.}\ \bibnamefont
  {Nguyen}}, \bibinfo {author} {\bibfnamefont {D.}~\bibnamefont {Timmer}},
  \bibinfo {author} {\bibfnamefont {Y.}~\bibnamefont {Rakita}}, \bibinfo
  {author} {\bibfnamefont {D.}~\bibnamefont {Cahen}}, \bibinfo {author}
  {\bibfnamefont {A.}~\bibnamefont {Steinhoff}}, \bibinfo {author}
  {\bibfnamefont {F.}~\bibnamefont {Jahnke}}, \bibinfo {author} {\bibfnamefont
  {C.}~\bibnamefont {Lienau}}, \ and\ \bibinfo {author} {\bibfnamefont
  {A.}~\bibnamefont {De~Sio}},\ }\href {\doibase 10.1021/acs.jpclett.9b01936}
  {\bibfield  {journal} {\bibinfo  {journal} {J. Phys. Chem. Lett.}\ }\textbf
  {\bibinfo {volume} {10}},\ \bibinfo {pages} {5414} (\bibinfo {year}
  {2019})}\BibitemShut {NoStop}%
\bibitem [{\citenamefont {Li}\ \emph {et~al.}(2006)\citenamefont {Li},
  \citenamefont {Zhang}, \citenamefont {Borca},\ and\ \citenamefont
  {Cundiff}}]{Li2006}%
  \BibitemOpen
  \bibfield  {author} {\bibinfo {author} {\bibfnamefont {X.}~\bibnamefont
  {Li}}, \bibinfo {author} {\bibfnamefont {T.}~\bibnamefont {Zhang}}, \bibinfo
  {author} {\bibfnamefont {C.~N.}\ \bibnamefont {Borca}}, \ and\ \bibinfo
  {author} {\bibfnamefont {S.~T.}\ \bibnamefont {Cundiff}},\ }\href {\doibase
  10.1103/PhysRevLett.96.057406} {\bibfield  {journal} {\bibinfo  {journal}
  {Phys. Rev. Lett.}\ }\textbf {\bibinfo {volume} {96}},\ \bibinfo {pages}
  {057406} (\bibinfo {year} {2006})}\BibitemShut {NoStop}%
\bibitem [{\citenamefont {Moody}\ \emph {et~al.}(2011)\citenamefont {Moody},
  \citenamefont {Siemens}, \citenamefont {Bristow}, \citenamefont {Dai},
  \citenamefont {Karaiskaj}, \citenamefont {Bracker}, \citenamefont {Gammon},\
  and\ \citenamefont {Cundiff}}]{Moody2011}%
  \BibitemOpen
  \bibfield  {author} {\bibinfo {author} {\bibfnamefont {G.}~\bibnamefont
  {Moody}}, \bibinfo {author} {\bibfnamefont {M.~E.}\ \bibnamefont {Siemens}},
  \bibinfo {author} {\bibfnamefont {A.~D.}\ \bibnamefont {Bristow}}, \bibinfo
  {author} {\bibfnamefont {X.}~\bibnamefont {Dai}}, \bibinfo {author}
  {\bibfnamefont {D.}~\bibnamefont {Karaiskaj}}, \bibinfo {author}
  {\bibfnamefont {A.~S.}\ \bibnamefont {Bracker}}, \bibinfo {author}
  {\bibfnamefont {D.}~\bibnamefont {Gammon}}, \ and\ \bibinfo {author}
  {\bibfnamefont {S.~T.}\ \bibnamefont {Cundiff}},\ }\href {\doibase
  10.1103/PhysRevB.83.115324} {\bibfield  {journal} {\bibinfo  {journal} {Phys.
  Rev. B}\ }\textbf {\bibinfo {volume} {83}},\ \bibinfo {pages} {115324}
  (\bibinfo {year} {2011})}\BibitemShut {NoStop}%
\bibitem [{\citenamefont {Stone}\ \emph {et~al.}(2009)\citenamefont {Stone},
  \citenamefont {Turner}, \citenamefont {Gundogdu}, \citenamefont {Cundiff},\
  and\ \citenamefont {Nelson}}]{Stone2009}%
  \BibitemOpen
  \bibfield  {author} {\bibinfo {author} {\bibfnamefont {K.~W.}\ \bibnamefont
  {Stone}}, \bibinfo {author} {\bibfnamefont {D.~B.}\ \bibnamefont {Turner}},
  \bibinfo {author} {\bibfnamefont {K.}~\bibnamefont {Gundogdu}}, \bibinfo
  {author} {\bibfnamefont {S.~T.}\ \bibnamefont {Cundiff}}, \ and\ \bibinfo
  {author} {\bibfnamefont {K.~A.}\ \bibnamefont {Nelson}},\ }\href {\doibase
  10.1021/ar900122k} {\bibfield  {journal} {\bibinfo  {journal} {Acc. Chem.
  Res}\ }\textbf {\bibinfo {volume} {42}},\ \bibinfo {pages} {1452} (\bibinfo
  {year} {2009})}\BibitemShut {NoStop}%
\bibitem [{\citenamefont {Turner}\ and\ \citenamefont
  {Nelson}(2010)}]{Turner2010}%
  \BibitemOpen
  \bibfield  {author} {\bibinfo {author} {\bibfnamefont {D.~B.}\ \bibnamefont
  {Turner}}\ and\ \bibinfo {author} {\bibfnamefont {K.~A.}\ \bibnamefont
  {Nelson}},\ }\href {\doibase 10.1038/nature09286} {\bibfield  {journal}
  {\bibinfo  {journal} {Nature}\ }\textbf {\bibinfo {volume} {466}},\ \bibinfo
  {pages} {1089} (\bibinfo {year} {2010})}\BibitemShut {NoStop}%
\bibitem [{\citenamefont {Collini}\ \emph {et~al.}(2020)\citenamefont
  {Collini}, \citenamefont {Gattuso}, \citenamefont {Kolodny}, \citenamefont
  {Bolzonello}, \citenamefont {Volpato}, \citenamefont {Fridman}, \citenamefont
  {Yochelis}, \citenamefont {Mor}, \citenamefont {Dehnel}, \citenamefont
  {Lifshitz}, \citenamefont {Paltiel}, \citenamefont {Levine},\ and\
  \citenamefont {Remacle}}]{Collini2020}%
  \BibitemOpen
  \bibfield  {author} {\bibinfo {author} {\bibfnamefont {E.}~\bibnamefont
  {Collini}}, \bibinfo {author} {\bibfnamefont {H.}~\bibnamefont {Gattuso}},
  \bibinfo {author} {\bibfnamefont {Y.}~\bibnamefont {Kolodny}}, \bibinfo
  {author} {\bibfnamefont {L.}~\bibnamefont {Bolzonello}}, \bibinfo {author}
  {\bibfnamefont {A.}~\bibnamefont {Volpato}}, \bibinfo {author} {\bibfnamefont
  {H.~T.}\ \bibnamefont {Fridman}}, \bibinfo {author} {\bibfnamefont
  {S.}~\bibnamefont {Yochelis}}, \bibinfo {author} {\bibfnamefont
  {M.}~\bibnamefont {Mor}}, \bibinfo {author} {\bibfnamefont {J.}~\bibnamefont
  {Dehnel}}, \bibinfo {author} {\bibfnamefont {E.}~\bibnamefont {Lifshitz}},
  \bibinfo {author} {\bibfnamefont {Y.}~\bibnamefont {Paltiel}}, \bibinfo
  {author} {\bibfnamefont {R.~D.}\ \bibnamefont {Levine}}, \ and\ \bibinfo
  {author} {\bibfnamefont {F.}~\bibnamefont {Remacle}},\ }\href {\doibase
  10.1021/acs.jpcc.0c05572} {\bibfield  {journal} {\bibinfo  {journal} {J.
  Phys. Chem. C}\ }\textbf {\bibinfo {volume} {124}},\ \bibinfo {pages} {16222}
  (\bibinfo {year} {2020})}\BibitemShut {NoStop}%
\bibitem [{\citenamefont {Collini}\ \emph {et~al.}(2021)\citenamefont
  {Collini}, \citenamefont {Gattuso}, \citenamefont {Levine},\ and\
  \citenamefont {Remacle}}]{Collini2021}%
  \BibitemOpen
  \bibfield  {author} {\bibinfo {author} {\bibfnamefont {E.}~\bibnamefont
  {Collini}}, \bibinfo {author} {\bibfnamefont {H.}~\bibnamefont {Gattuso}},
  \bibinfo {author} {\bibfnamefont {R.~D.}\ \bibnamefont {Levine}}, \ and\
  \bibinfo {author} {\bibfnamefont {F.}~\bibnamefont {Remacle}},\ }\href
  {\doibase 10.1063/5.0031420} {\bibfield  {journal} {\bibinfo  {journal} {J.
  Chem. Phys.}\ }\textbf {\bibinfo {volume} {154}},\ \bibinfo {pages} {014301}
  (\bibinfo {year} {2021})}\BibitemShut {NoStop}%
\bibitem [{\citenamefont {Liu}\ \emph {et~al.}(2022)\citenamefont {Liu},
  \citenamefont {Almeida}, \citenamefont {Padilha},\ and\ \citenamefont
  {Cundiff}}]{Liu_2022}%
  \BibitemOpen
  \bibfield  {author} {\bibinfo {author} {\bibfnamefont {A.}~\bibnamefont
  {Liu}}, \bibinfo {author} {\bibfnamefont {D.~B.}\ \bibnamefont {Almeida}},
  \bibinfo {author} {\bibfnamefont {L.~A.}\ \bibnamefont {Padilha}}, \ and\
  \bibinfo {author} {\bibfnamefont {S.~T.}\ \bibnamefont {Cundiff}},\ }\href
  {\doibase 10.1088/2515-7639/ac4fa5} {\bibfield  {journal} {\bibinfo
  {journal} {J. Phys. Materials}\ }\textbf {\bibinfo {volume} {5}},\ \bibinfo
  {pages} {021002} (\bibinfo {year} {2022})}\BibitemShut {NoStop}%
\bibitem [{\citenamefont {Srimath~Kandada}\ \emph {et~al.}(2020)\citenamefont
  {Srimath~Kandada}, \citenamefont {Li}, \citenamefont {Thouin}, \citenamefont
  {Bittner},\ and\ \citenamefont {Silva}}]{Kandada2020Stochastic}%
  \BibitemOpen
  \bibfield  {author} {\bibinfo {author} {\bibfnamefont {A.~R.}\ \bibnamefont
  {Srimath~Kandada}}, \bibinfo {author} {\bibfnamefont {H.}~\bibnamefont {Li}},
  \bibinfo {author} {\bibfnamefont {F.}~\bibnamefont {Thouin}}, \bibinfo
  {author} {\bibfnamefont {E.~R.}\ \bibnamefont {Bittner}}, \ and\ \bibinfo
  {author} {\bibfnamefont {C.}~\bibnamefont {Silva}},\ }\href {\doibase
  10.1063/5.0026351} {\bibfield  {journal} {\bibinfo  {journal} {J. Chem.
  Phys.}\ }\textbf {\bibinfo {volume} {153}},\ \bibinfo {pages} {164706}
  (\bibinfo {year} {2020})}\BibitemShut {NoStop}%
\bibitem [{\citenamefont {Allerbeck}\ \emph {et~al.}(2021)\citenamefont
  {Allerbeck}, \citenamefont {Deckert}, \citenamefont {Spitzner},\ and\
  \citenamefont {Brida}}]{Allerber2021}%
  \BibitemOpen
  \bibfield  {author} {\bibinfo {author} {\bibfnamefont {J.}~\bibnamefont
  {Allerbeck}}, \bibinfo {author} {\bibfnamefont {T.}~\bibnamefont {Deckert}},
  \bibinfo {author} {\bibfnamefont {L.}~\bibnamefont {Spitzner}}, \ and\
  \bibinfo {author} {\bibfnamefont {D.}~\bibnamefont {Brida}},\ }\href
  {\doibase 10.1103/PhysRevB.104.L201202} {\bibfield  {journal} {\bibinfo
  {journal} {Phys. Rev. B}\ }\textbf {\bibinfo {volume} {104}},\ \bibinfo
  {pages} {L201202} (\bibinfo {year} {2021})}\BibitemShut {NoStop}%
\bibitem [{\citenamefont {Rojas-Gatjens}\ \emph {et~al.}(2023)\citenamefont
  {Rojas-Gatjens}, \citenamefont {Li}, \citenamefont {Vega-Flick},
  \citenamefont {Cortecchia}, \citenamefont {Petrozza}, \citenamefont
  {Bittner}, \citenamefont {Srimath~Kandada},\ and\ \citenamefont
  {Silva-Acuña}}]{Rojas2023manyexciton}%
  \BibitemOpen
  \bibfield  {author} {\bibinfo {author} {\bibfnamefont {E.}~\bibnamefont
  {Rojas-Gatjens}}, \bibinfo {author} {\bibfnamefont {H.}~\bibnamefont {Li}},
  \bibinfo {author} {\bibfnamefont {A.}~\bibnamefont {Vega-Flick}}, \bibinfo
  {author} {\bibfnamefont {D.}~\bibnamefont {Cortecchia}}, \bibinfo {author}
  {\bibfnamefont {A.}~\bibnamefont {Petrozza}}, \bibinfo {author}
  {\bibfnamefont {E.~R.}\ \bibnamefont {Bittner}}, \bibinfo {author}
  {\bibfnamefont {A.~R.}\ \bibnamefont {Srimath~Kandada}}, \ and\ \bibinfo
  {author} {\bibfnamefont {C.}~\bibnamefont {Silva-Acuña}},\ }\href@noop {}
  {\enquote {\bibinfo {title} {Many-exciton quantum dynamics in a
  ruddlesden-popper tin iodide},}\ } (\bibinfo {year} {2023}),\ \Eprint
  {http://arxiv.org/abs/2304.02461} {arXiv:2304.02461 [cond-mat.mtrl-sci]}
  \BibitemShut {NoStop}%
\bibitem [{\citenamefont {Richter}\ \emph {et~al.}(2017)\citenamefont
  {Richter}, \citenamefont {Branchi}, \citenamefont {Valduga~de
  Almeida~Camargo}, \citenamefont {Zhao}, \citenamefont {Friend}, \citenamefont
  {Cerullo},\ and\ \citenamefont {Deschler}}]{Richter2017}%
  \BibitemOpen
  \bibfield  {author} {\bibinfo {author} {\bibfnamefont {J.~M.}\ \bibnamefont
  {Richter}}, \bibinfo {author} {\bibfnamefont {F.}~\bibnamefont {Branchi}},
  \bibinfo {author} {\bibfnamefont {F.}~\bibnamefont {Valduga~de
  Almeida~Camargo}}, \bibinfo {author} {\bibfnamefont {B.}~\bibnamefont
  {Zhao}}, \bibinfo {author} {\bibfnamefont {R.~H.}\ \bibnamefont {Friend}},
  \bibinfo {author} {\bibfnamefont {G.}~\bibnamefont {Cerullo}}, \ and\
  \bibinfo {author} {\bibfnamefont {F.}~\bibnamefont {Deschler}},\ }\href@noop
  {} {\bibfield  {journal} {\bibinfo  {journal} {Nat. Commun.}\ }\textbf
  {\bibinfo {volume} {8}},\ \bibinfo {pages} {376} (\bibinfo {year}
  {2017})}\BibitemShut {NoStop}%
\bibitem [{\citenamefont {Yu}\ \emph {et~al.}(2021)\citenamefont {Yu},
  \citenamefont {Chen}, \citenamefont {Qu}, \citenamefont {Zhang},
  \citenamefont {Qin}, \citenamefont {Wang},\ and\ \citenamefont
  {Xiao}}]{Yu2021}%
  \BibitemOpen
  \bibfield  {author} {\bibinfo {author} {\bibfnamefont {B.}~\bibnamefont
  {Yu}}, \bibinfo {author} {\bibfnamefont {L.}~\bibnamefont {Chen}}, \bibinfo
  {author} {\bibfnamefont {Z.}~\bibnamefont {Qu}}, \bibinfo {author}
  {\bibfnamefont {C.}~\bibnamefont {Zhang}}, \bibinfo {author} {\bibfnamefont
  {Z.}~\bibnamefont {Qin}}, \bibinfo {author} {\bibfnamefont {X.}~\bibnamefont
  {Wang}}, \ and\ \bibinfo {author} {\bibfnamefont {M.}~\bibnamefont {Xiao}},\
  }\href {\doibase 10.1021/acs.jpclett.0c03350} {\bibfield  {journal} {\bibinfo
   {journal} {J. Phys. Chem. Lett.}\ }\textbf {\bibinfo {volume} {12}},\
  \bibinfo {pages} {238} (\bibinfo {year} {2021})}\BibitemShut {NoStop}%
\bibitem [{\citenamefont {Turner}\ \emph {et~al.}(2011)\citenamefont {Turner},
  \citenamefont {Stone}, \citenamefont {Gundogdu},\ and\ \citenamefont
  {Nelson}}]{Turner2011}%
  \BibitemOpen
  \bibfield  {author} {\bibinfo {author} {\bibfnamefont {D.~B.}\ \bibnamefont
  {Turner}}, \bibinfo {author} {\bibfnamefont {K.~W.}\ \bibnamefont {Stone}},
  \bibinfo {author} {\bibfnamefont {K.}~\bibnamefont {Gundogdu}}, \ and\
  \bibinfo {author} {\bibfnamefont {K.~A.}\ \bibnamefont {Nelson}},\ }\href
  {\doibase 10.1063/1.3624752} {\bibfield  {journal} {\bibinfo  {journal} {Rev.
  Sci. Instrum}\ }\textbf {\bibinfo {volume} {82}} (\bibinfo {year} {2011}),\
  10.1063/1.3624752}\BibitemShut {NoStop}%
\bibitem [{\citenamefont {Romero-Pérez}\ \emph {et~al.}(2023)\citenamefont
  {Romero-Pérez}, \citenamefont {Delgado}, \citenamefont {Herrera-Collado},
  \citenamefont {Calvo},\ and\ \citenamefont {Míguez}}]{romero2023}%
  \BibitemOpen
  \bibfield  {author} {\bibinfo {author} {\bibfnamefont {C.}~\bibnamefont
  {Romero-Pérez}}, \bibinfo {author} {\bibfnamefont {N.~F.}\ \bibnamefont
  {Delgado}}, \bibinfo {author} {\bibfnamefont {M.}~\bibnamefont
  {Herrera-Collado}}, \bibinfo {author} {\bibfnamefont {M.~E.}\ \bibnamefont
  {Calvo}}, \ and\ \bibinfo {author} {\bibfnamefont {H.}~\bibnamefont
  {Míguez}},\ }\href {\doibase 10.1021/acs.chemmater.3c00934} {\bibfield
  {journal} {\bibinfo  {journal} {Chem. Mat.}\ }\textbf {\bibinfo {volume}
  {35}},\ \bibinfo {pages} {5541} (\bibinfo {year} {2023})}\BibitemShut
  {NoStop}%
\bibitem [{\citenamefont {Perumal}\ \emph {et~al.}(2016)\citenamefont
  {Perumal}, \citenamefont {Li}, \citenamefont {Tay}, \citenamefont {Sharma},
  \citenamefont {Chen}, \citenamefont {Wei}, \citenamefont {Liu}, \citenamefont
  {Gao}, \citenamefont {Buenconsejo}, \citenamefont {Tan}, \citenamefont {Gan},
  \citenamefont {Xiong}, \citenamefont {Sum},\ and\ \citenamefont
  {Demir}}]{Perumal2016}%
  \BibitemOpen
  \bibfield  {author} {\bibinfo {author} {\bibfnamefont {S.}~\bibnamefont
  {Perumal}, \bibfnamefont {Ajayand~Shendre}}, \bibinfo {author} {\bibfnamefont
  {M.}~\bibnamefont {Li}}, \bibinfo {author} {\bibfnamefont {Y.~K.~E.}\
  \bibnamefont {Tay}}, \bibinfo {author} {\bibfnamefont {V.~K.}\ \bibnamefont
  {Sharma}}, \bibinfo {author} {\bibfnamefont {S.}~\bibnamefont {Chen}},
  \bibinfo {author} {\bibfnamefont {Z.}~\bibnamefont {Wei}}, \bibinfo {author}
  {\bibfnamefont {Q.}~\bibnamefont {Liu}}, \bibinfo {author} {\bibfnamefont
  {Y.}~\bibnamefont {Gao}}, \bibinfo {author} {\bibfnamefont {P.~J.~S.}\
  \bibnamefont {Buenconsejo}}, \bibinfo {author} {\bibfnamefont {S.~T.}\
  \bibnamefont {Tan}}, \bibinfo {author} {\bibfnamefont {C.~L.}\ \bibnamefont
  {Gan}}, \bibinfo {author} {\bibfnamefont {Q.}~\bibnamefont {Xiong}}, \bibinfo
  {author} {\bibfnamefont {T.~C.}\ \bibnamefont {Sum}}, \ and\ \bibinfo
  {author} {\bibfnamefont {H.~V.}\ \bibnamefont {Demir}},\ }\href {\doibase
  10.1038/srep36733} {\bibfield  {journal} {\bibinfo  {journal} {Sci. Rep.}\
  }\textbf {\bibinfo {volume} {6}},\ \bibinfo {pages} {36733} (\bibinfo {year}
  {2016})}\BibitemShut {NoStop}%
\bibitem [{\citenamefont {Baumgardner}\ \emph {et~al.}(2013)\citenamefont
  {Baumgardner}, \citenamefont {Whitham},\ and\ \citenamefont
  {Hanrath}}]{Baumgardner2013}%
  \BibitemOpen
  \bibfield  {author} {\bibinfo {author} {\bibfnamefont {W.~J.}\ \bibnamefont
  {Baumgardner}}, \bibinfo {author} {\bibfnamefont {K.}~\bibnamefont
  {Whitham}}, \ and\ \bibinfo {author} {\bibfnamefont {T.}~\bibnamefont
  {Hanrath}},\ }\href {\doibase 10.1021/nl401298s} {\bibfield  {journal}
  {\bibinfo  {journal} {Nano Letters}\ }\textbf {\bibinfo {volume} {13}},\
  \bibinfo {pages} {3225} (\bibinfo {year} {2013})},\ \bibinfo {note} {pMID:
  23777454}\BibitemShut {NoStop}%
\bibitem [{\citenamefont {Yang}\ \emph {et~al.}(2007)\citenamefont {Yang},
  \citenamefont {Schweigert}, \citenamefont {Cundiff},\ and\ \citenamefont
  {Mukamel}}]{Yang2007Correlations}%
  \BibitemOpen
  \bibfield  {author} {\bibinfo {author} {\bibfnamefont {L.}~\bibnamefont
  {Yang}}, \bibinfo {author} {\bibfnamefont {I.~V.}\ \bibnamefont
  {Schweigert}}, \bibinfo {author} {\bibfnamefont {S.~T.}\ \bibnamefont
  {Cundiff}}, \ and\ \bibinfo {author} {\bibfnamefont {S.}~\bibnamefont
  {Mukamel}},\ }\href {\doibase 10.1103/PhysRevB.75.125302} {\bibfield
  {journal} {\bibinfo  {journal} {Phys. Rev. B}\ }\textbf {\bibinfo {volume}
  {75}},\ \bibinfo {pages} {125302} (\bibinfo {year} {2007})}\BibitemShut
  {NoStop}%
\bibitem [{\citenamefont {Yang}\ \emph {et~al.}(2008)\citenamefont {Yang},
  \citenamefont {Zhang}, \citenamefont {Bristow}, \citenamefont {Cundiff},\
  and\ \citenamefont {Mukamel}}]{Yang2008Isolating}%
  \BibitemOpen
  \bibfield  {author} {\bibinfo {author} {\bibfnamefont {L.}~\bibnamefont
  {Yang}}, \bibinfo {author} {\bibfnamefont {T.}~\bibnamefont {Zhang}},
  \bibinfo {author} {\bibfnamefont {A.~D.}\ \bibnamefont {Bristow}}, \bibinfo
  {author} {\bibfnamefont {S.~T.}\ \bibnamefont {Cundiff}}, \ and\ \bibinfo
  {author} {\bibfnamefont {S.}~\bibnamefont {Mukamel}},\ }\href {\doibase
  10.1063/1.3037217} {\bibfield  {journal} {\bibinfo  {journal} {J. Chem.
  Phys.}\ }\textbf {\bibinfo {volume} {129}},\ \bibinfo {pages} {234711}
  (\bibinfo {year} {2008})}\BibitemShut {NoStop}%
\bibitem [{\citenamefont {Huang}\ \emph {et~al.}(2020)\citenamefont {Huang},
  \citenamefont {Chen}, \citenamefont {Zhang}, \citenamefont {Qin},
  \citenamefont {Yu}, \citenamefont {Wang},\ and\ \citenamefont
  {Xiao}}]{Huang2020}%
  \BibitemOpen
  \bibfield  {author} {\bibinfo {author} {\bibfnamefont {X.}~\bibnamefont
  {Huang}}, \bibinfo {author} {\bibfnamefont {L.}~\bibnamefont {Chen}},
  \bibinfo {author} {\bibfnamefont {C.}~\bibnamefont {Zhang}}, \bibinfo
  {author} {\bibfnamefont {Z.}~\bibnamefont {Qin}}, \bibinfo {author}
  {\bibfnamefont {B.}~\bibnamefont {Yu}}, \bibinfo {author} {\bibfnamefont
  {X.}~\bibnamefont {Wang}}, \ and\ \bibinfo {author} {\bibfnamefont
  {M.}~\bibnamefont {Xiao}},\ }\href {\doibase 10.1021/acs.jpclett.0c03153}
  {\bibfield  {journal} {\bibinfo  {journal} {J. Phys. Chem. Lett.}\ }\textbf
  {\bibinfo {volume} {11}},\ \bibinfo {pages} {10173} (\bibinfo {year}
  {2020})}\BibitemShut {NoStop}%
\bibitem [{\citenamefont {Zhao}\ \emph {et~al.}(2019)\citenamefont {Zhao},
  \citenamefont {Qin}, \citenamefont {Zhang}, \citenamefont {Wang},
  \citenamefont {Dai},\ and\ \citenamefont {Xiao}}]{Zhao2019}%
  \BibitemOpen
  \bibfield  {author} {\bibinfo {author} {\bibfnamefont {W.}~\bibnamefont
  {Zhao}}, \bibinfo {author} {\bibfnamefont {Z.}~\bibnamefont {Qin}}, \bibinfo
  {author} {\bibfnamefont {C.}~\bibnamefont {Zhang}}, \bibinfo {author}
  {\bibfnamefont {G.}~\bibnamefont {Wang}}, \bibinfo {author} {\bibfnamefont
  {X.}~\bibnamefont {Dai}}, \ and\ \bibinfo {author} {\bibfnamefont
  {M.}~\bibnamefont {Xiao}},\ }\href {\doibase 10.1063/1.5130636} {\bibfield
  {journal} {\bibinfo  {journal} {Appl. Phys. Lett.}\ }\textbf {\bibinfo
  {volume} {115}} (\bibinfo {year} {2019}),\ 10.1063/1.5130636}\BibitemShut
  {NoStop}%
\bibitem [{\citenamefont {Shacklette}\ and\ \citenamefont
  {Cundiff}(2002)}]{Shacklette2002}%
  \BibitemOpen
  \bibfield  {author} {\bibinfo {author} {\bibfnamefont {J.~M.}\ \bibnamefont
  {Shacklette}}\ and\ \bibinfo {author} {\bibfnamefont {S.~T.}\ \bibnamefont
  {Cundiff}},\ }\href {\doibase 10.1103/PhysRevB.66.045309} {\bibfield
  {journal} {\bibinfo  {journal} {Phys. Rev. B}\ }\textbf {\bibinfo {volume}
  {66}},\ \bibinfo {pages} {045309} (\bibinfo {year} {2002})}\BibitemShut
  {NoStop}%
\bibitem [{\citenamefont {Shacklette}\ and\ \citenamefont
  {Cundiff}(2003)}]{Shacklette2003}%
  \BibitemOpen
  \bibfield  {author} {\bibinfo {author} {\bibfnamefont {J.~M.}\ \bibnamefont
  {Shacklette}}\ and\ \bibinfo {author} {\bibfnamefont {S.~T.}\ \bibnamefont
  {Cundiff}},\ }\href {\doibase 10.1364/JOSAB.20.000764} {\bibfield  {journal}
  {\bibinfo  {journal} {J. Opt. Soc. Am. B}\ }\textbf {\bibinfo {volume}
  {20}},\ \bibinfo {pages} {764} (\bibinfo {year} {2003})}\BibitemShut
  {NoStop}%
\bibitem [{\citenamefont {Bristow}\ \emph {et~al.}(2009)\citenamefont
  {Bristow}, \citenamefont {Karaiskaj}, \citenamefont {Dai}, \citenamefont
  {Mirin},\ and\ \citenamefont {Cundiff}}]{Bristow2009}%
  \BibitemOpen
  \bibfield  {author} {\bibinfo {author} {\bibfnamefont {A.~D.}\ \bibnamefont
  {Bristow}}, \bibinfo {author} {\bibfnamefont {D.}~\bibnamefont {Karaiskaj}},
  \bibinfo {author} {\bibfnamefont {X.}~\bibnamefont {Dai}}, \bibinfo {author}
  {\bibfnamefont {R.~P.}\ \bibnamefont {Mirin}}, \ and\ \bibinfo {author}
  {\bibfnamefont {S.~T.}\ \bibnamefont {Cundiff}},\ }\href {\doibase
  10.1103/PhysRevB.79.161305} {\bibfield  {journal} {\bibinfo  {journal} {Phys.
  Rev. B}\ }\textbf {\bibinfo {volume} {79}},\ \bibinfo {pages} {161305}
  (\bibinfo {year} {2009})}\BibitemShut {NoStop}%
\bibitem [{\citenamefont {Li}\ \emph {et~al.}(2020)\citenamefont {Li},
  \citenamefont {Srimath~Kandada}, \citenamefont {Silva},\ and\ \citenamefont
  {Bittner}}]{Li2020}%
  \BibitemOpen
  \bibfield  {author} {\bibinfo {author} {\bibfnamefont {H.}~\bibnamefont
  {Li}}, \bibinfo {author} {\bibfnamefont {A.~R.}\ \bibnamefont
  {Srimath~Kandada}}, \bibinfo {author} {\bibfnamefont {C.}~\bibnamefont
  {Silva}}, \ and\ \bibinfo {author} {\bibfnamefont {E.~R.}\ \bibnamefont
  {Bittner}},\ }\href {\doibase 10.1063/5.0026467} {\bibfield  {journal}
  {\bibinfo  {journal} {J. Chem. Phys.}\ }\textbf {\bibinfo {volume} {153}},\
  \bibinfo {pages} {154115} (\bibinfo {year} {2020})}\BibitemShut {NoStop}%
\bibitem [{\citenamefont {Li}\ \emph {et~al.}(2023{\natexlab{a}})\citenamefont
  {Li}, \citenamefont {Shah}, \citenamefont {Kandada}, \citenamefont {Silva},
  \citenamefont {Piryatinski},\ and\ \citenamefont {Bittner}}]{Li2023}%
  \BibitemOpen
  \bibfield  {author} {\bibinfo {author} {\bibfnamefont {H.}~\bibnamefont
  {Li}}, \bibinfo {author} {\bibfnamefont {S.}~\bibnamefont {Shah}}, \bibinfo
  {author} {\bibfnamefont {A.~R.~S.}\ \bibnamefont {Kandada}}, \bibinfo
  {author} {\bibfnamefont {C.}~\bibnamefont {Silva}}, \bibinfo {author}
  {\bibfnamefont {A.}~\bibnamefont {Piryatinski}}, \ and\ \bibinfo {author}
  {\bibfnamefont {E.~R.}\ \bibnamefont {Bittner}},\ }\href {\doibase
  10.1146/annurev-physchem-102822-100922} {\bibfield  {journal} {\bibinfo
  {journal} {Annu. Rev. Phys. Chem.}\ }\textbf {\bibinfo {volume} {74}},\
  \bibinfo {pages} {467} (\bibinfo {year} {2023}{\natexlab{a}})}\BibitemShut
  {NoStop}%
\bibitem [{\citenamefont {Trovatello}\ \emph {et~al.}(2022)\citenamefont
  {Trovatello}, \citenamefont {Katsch}, \citenamefont {Li}, \citenamefont
  {Zhu}, \citenamefont {Knorr}, \citenamefont {Cerullo},\ and\ \citenamefont
  {Dal~Conte}}]{Trovatello2022}%
  \BibitemOpen
  \bibfield  {author} {\bibinfo {author} {\bibfnamefont {C.}~\bibnamefont
  {Trovatello}}, \bibinfo {author} {\bibfnamefont {F.}~\bibnamefont {Katsch}},
  \bibinfo {author} {\bibfnamefont {Q.}~\bibnamefont {Li}}, \bibinfo {author}
  {\bibfnamefont {X.}~\bibnamefont {Zhu}}, \bibinfo {author} {\bibfnamefont
  {A.}~\bibnamefont {Knorr}}, \bibinfo {author} {\bibfnamefont
  {G.}~\bibnamefont {Cerullo}}, \ and\ \bibinfo {author} {\bibfnamefont
  {S.}~\bibnamefont {Dal~Conte}},\ }\href {\doibase
  10.1021/acs.nanolett.2c01309} {\bibfield  {journal} {\bibinfo  {journal}
  {Nano Lett.}\ }\textbf {\bibinfo {volume} {22}},\ \bibinfo {pages} {5322}
  (\bibinfo {year} {2022})}\BibitemShut {NoStop}%
\bibitem [{\citenamefont {Baiz}\ \emph {et~al.}(2020)\citenamefont {Baiz},
  \citenamefont {Błasiak}, \citenamefont {Bredenbeck}, \citenamefont {Cho},
  \citenamefont {Choi}, \citenamefont {Corcelli}, \citenamefont {Dijkstra},
  \citenamefont {Feng}, \citenamefont {Garrett-Roe}, \citenamefont {Ge},
  \citenamefont {Hanson-Heine}, \citenamefont {Hirst}, \citenamefont {Jansen},
  \citenamefont {Kwac}, \citenamefont {Kubarych}, \citenamefont {Londergan},
  \citenamefont {Maekawa}, \citenamefont {Reppert}, \citenamefont {Saito},
  \citenamefont {Roy}, \citenamefont {Skinner}, \citenamefont {Stock},
  \citenamefont {Straub}, \citenamefont {Thielges}, \citenamefont {Tominaga},
  \citenamefont {Tokmakoff}, \citenamefont {Torii}, \citenamefont {Wang},
  \citenamefont {Webb},\ and\ \citenamefont {Zanni}}]{Baiz2020}%
  \BibitemOpen
  \bibfield  {author} {\bibinfo {author} {\bibfnamefont {C.~R.}\ \bibnamefont
  {Baiz}}, \bibinfo {author} {\bibfnamefont {B.}~\bibnamefont {Błasiak}},
  \bibinfo {author} {\bibfnamefont {J.}~\bibnamefont {Bredenbeck}}, \bibinfo
  {author} {\bibfnamefont {M.}~\bibnamefont {Cho}}, \bibinfo {author}
  {\bibfnamefont {J.-H.}\ \bibnamefont {Choi}}, \bibinfo {author}
  {\bibfnamefont {S.~A.}\ \bibnamefont {Corcelli}}, \bibinfo {author}
  {\bibfnamefont {A.~G.}\ \bibnamefont {Dijkstra}}, \bibinfo {author}
  {\bibfnamefont {C.-J.}\ \bibnamefont {Feng}}, \bibinfo {author}
  {\bibfnamefont {S.}~\bibnamefont {Garrett-Roe}}, \bibinfo {author}
  {\bibfnamefont {N.-H.}\ \bibnamefont {Ge}}, \bibinfo {author} {\bibfnamefont
  {M.~W.~D.}\ \bibnamefont {Hanson-Heine}}, \bibinfo {author} {\bibfnamefont
  {J.~D.}\ \bibnamefont {Hirst}}, \bibinfo {author} {\bibfnamefont {T.~L.~C.}\
  \bibnamefont {Jansen}}, \bibinfo {author} {\bibfnamefont {K.}~\bibnamefont
  {Kwac}}, \bibinfo {author} {\bibfnamefont {K.~J.}\ \bibnamefont {Kubarych}},
  \bibinfo {author} {\bibfnamefont {C.~H.}\ \bibnamefont {Londergan}}, \bibinfo
  {author} {\bibfnamefont {H.}~\bibnamefont {Maekawa}}, \bibinfo {author}
  {\bibfnamefont {M.}~\bibnamefont {Reppert}}, \bibinfo {author} {\bibfnamefont
  {S.}~\bibnamefont {Saito}}, \bibinfo {author} {\bibfnamefont
  {S.}~\bibnamefont {Roy}}, \bibinfo {author} {\bibfnamefont {J.~L.}\
  \bibnamefont {Skinner}}, \bibinfo {author} {\bibfnamefont {G.}~\bibnamefont
  {Stock}}, \bibinfo {author} {\bibfnamefont {J.~E.}\ \bibnamefont {Straub}},
  \bibinfo {author} {\bibfnamefont {M.~C.}\ \bibnamefont {Thielges}}, \bibinfo
  {author} {\bibfnamefont {K.}~\bibnamefont {Tominaga}}, \bibinfo {author}
  {\bibfnamefont {A.}~\bibnamefont {Tokmakoff}}, \bibinfo {author}
  {\bibfnamefont {H.}~\bibnamefont {Torii}}, \bibinfo {author} {\bibfnamefont
  {L.}~\bibnamefont {Wang}}, \bibinfo {author} {\bibfnamefont {L.~J.}\
  \bibnamefont {Webb}}, \ and\ \bibinfo {author} {\bibfnamefont {M.~T.}\
  \bibnamefont {Zanni}},\ }\href {\doibase 10.1021/acs.chemrev.9b00813}
  {\bibfield  {journal} {\bibinfo  {journal} {Chem. Rev.}\ }\textbf {\bibinfo
  {volume} {120}},\ \bibinfo {pages} {7152} (\bibinfo {year}
  {2020})}\BibitemShut {NoStop}%
\bibitem [{\citenamefont {Wilmer}\ \emph {et~al.}(2016)\citenamefont {Wilmer},
  \citenamefont {Webber}, \citenamefont {Ashley}, \citenamefont {Hall},\ and\
  \citenamefont {Bristow}}]{Wilmer2016}%
  \BibitemOpen
  \bibfield  {author} {\bibinfo {author} {\bibfnamefont {B.~L.}\ \bibnamefont
  {Wilmer}}, \bibinfo {author} {\bibfnamefont {D.}~\bibnamefont {Webber}},
  \bibinfo {author} {\bibfnamefont {J.~M.}\ \bibnamefont {Ashley}}, \bibinfo
  {author} {\bibfnamefont {K.~C.}\ \bibnamefont {Hall}}, \ and\ \bibinfo
  {author} {\bibfnamefont {A.~D.}\ \bibnamefont {Bristow}},\ }\href {\doibase
  10.1103/PhysRevB.94.075207} {\bibfield  {journal} {\bibinfo  {journal} {Phys.
  Rev. B}\ }\textbf {\bibinfo {volume} {94}},\ \bibinfo {pages} {075207}
  (\bibinfo {year} {2016})}\BibitemShut {NoStop}%
\bibitem [{\citenamefont {Fano}(1961)}]{Fano1961}%
  \BibitemOpen
  \bibfield  {author} {\bibinfo {author} {\bibfnamefont {U.}~\bibnamefont
  {Fano}},\ }\href {\doibase 10.1103/PhysRev.124.1866} {\bibfield  {journal}
  {\bibinfo  {journal} {Phys. Rev.}\ }\textbf {\bibinfo {volume} {124}},\
  \bibinfo {pages} {1866} (\bibinfo {year} {1961})}\BibitemShut {NoStop}%
\bibitem [{\citenamefont {Finkelstein-Shapiro}\ \emph
  {et~al.}(2016)\citenamefont {Finkelstein-Shapiro}, \citenamefont {Poulsen},
  \citenamefont {Pullerits},\ and\ \citenamefont {Hansen}}]{Fano1}%
  \BibitemOpen
  \bibfield  {author} {\bibinfo {author} {\bibfnamefont {D.}~\bibnamefont
  {Finkelstein-Shapiro}}, \bibinfo {author} {\bibfnamefont {F.}~\bibnamefont
  {Poulsen}}, \bibinfo {author} {\bibfnamefont {T.~o.}\ \bibnamefont
  {Pullerits}}, \ and\ \bibinfo {author} {\bibfnamefont {T.}~\bibnamefont
  {Hansen}},\ }\href {\doibase 10.1103/PhysRevB.94.205137} {\bibfield
  {journal} {\bibinfo  {journal} {Phys. Rev. B}\ }\textbf {\bibinfo {volume}
  {94}},\ \bibinfo {pages} {205137} (\bibinfo {year} {2016})}\BibitemShut
  {NoStop}%
\bibitem [{\citenamefont {Finkelstein-Shapiro}\ \emph
  {et~al.}(2018)\citenamefont {Finkelstein-Shapiro}, \citenamefont
  {Pullerits},\ and\ \citenamefont {Hansen}}]{Fano2}%
  \BibitemOpen
  \bibfield  {author} {\bibinfo {author} {\bibfnamefont {D.}~\bibnamefont
  {Finkelstein-Shapiro}}, \bibinfo {author} {\bibfnamefont {T.}~\bibnamefont
  {Pullerits}}, \ and\ \bibinfo {author} {\bibfnamefont {T.}~\bibnamefont
  {Hansen}},\ }\href {\doibase 10.1063/1.5019376} {\bibfield  {journal}
  {\bibinfo  {journal} {J. Chem. Phys.}\ }\textbf {\bibinfo {volume} {148}}
  (\bibinfo {year} {2018}),\ 10.1063/1.5019376}\BibitemShut {NoStop}%
\bibitem [{\citenamefont {Li}\ \emph {et~al.}(2023{\natexlab{b}})\citenamefont
  {Li}, \citenamefont {Huang},\ and\ \citenamefont {Zhong}}]{Li2023_2}%
  \BibitemOpen
  \bibfield  {author} {\bibinfo {author} {\bibfnamefont {M.}~\bibnamefont
  {Li}}, \bibinfo {author} {\bibfnamefont {P.}~\bibnamefont {Huang}}, \ and\
  \bibinfo {author} {\bibfnamefont {H.}~\bibnamefont {Zhong}},\ }\href
  {\doibase 10.1021/acs.jpclett.2c03525} {\bibfield  {journal} {\bibinfo
  {journal} {J. Phys. Chem. Lett.}\ }\textbf {\bibinfo {volume} {14}},\
  \bibinfo {pages} {1592} (\bibinfo {year} {2023}{\natexlab{b}})}\BibitemShut
  {NoStop}%
\bibitem [{\citenamefont {Gogh}\ \emph {et~al.}(2009)\citenamefont {Gogh},
  \citenamefont {Thomas}, \citenamefont {Kuznetsova}, \citenamefont {Meier},\
  and\ \citenamefont {Varga}}]{G2009}%
  \BibitemOpen
  \bibfield  {author} {\bibinfo {author} {\bibfnamefont {N.}~\bibnamefont
  {Gogh}}, \bibinfo {author} {\bibfnamefont {P.}~\bibnamefont {Thomas}},
  \bibinfo {author} {\bibfnamefont {I.}~\bibnamefont {Kuznetsova}}, \bibinfo
  {author} {\bibfnamefont {T.}~\bibnamefont {Meier}}, \ and\ \bibinfo {author}
  {\bibfnamefont {I.}~\bibnamefont {Varga}},\ }\href {\doibase
  https://doi.org/10.1002/andp.20095211219} {\bibfield  {journal} {\bibinfo
  {journal} {Ann. Phys.}\ }\textbf {\bibinfo {volume} {521}},\ \bibinfo {pages}
  {905} (\bibinfo {year} {2009})}\BibitemShut {NoStop}%
\bibitem [{\citenamefont {Tiguntseva}\ \emph {et~al.}(2018)\citenamefont
  {Tiguntseva}, \citenamefont {Baranov}, \citenamefont {Pushkarev},
  \citenamefont {Munkhbat}, \citenamefont {Komissarenko}, \citenamefont
  {Franckevičius}, \citenamefont {Zakhidov}, \citenamefont {Shegai},
  \citenamefont {Kivshar},\ and\ \citenamefont {Makarov}}]{Tiguntseva2018}%
  \BibitemOpen
  \bibfield  {author} {\bibinfo {author} {\bibfnamefont {E.~Y.}\ \bibnamefont
  {Tiguntseva}}, \bibinfo {author} {\bibfnamefont {D.~G.}\ \bibnamefont
  {Baranov}}, \bibinfo {author} {\bibfnamefont {A.~P.}\ \bibnamefont
  {Pushkarev}}, \bibinfo {author} {\bibfnamefont {B.}~\bibnamefont {Munkhbat}},
  \bibinfo {author} {\bibfnamefont {F.}~\bibnamefont {Komissarenko}}, \bibinfo
  {author} {\bibfnamefont {M.}~\bibnamefont {Franckevičius}}, \bibinfo
  {author} {\bibfnamefont {A.~A.}\ \bibnamefont {Zakhidov}}, \bibinfo {author}
  {\bibfnamefont {T.}~\bibnamefont {Shegai}}, \bibinfo {author} {\bibfnamefont
  {Y.~S.}\ \bibnamefont {Kivshar}}, \ and\ \bibinfo {author} {\bibfnamefont
  {S.~V.}\ \bibnamefont {Makarov}},\ }\href {\doibase
  10.1021/acs.nanolett.8b01912} {\bibfield  {journal} {\bibinfo  {journal}
  {Nano Lett.}\ }\textbf {\bibinfo {volume} {18}},\ \bibinfo {pages} {5522}
  (\bibinfo {year} {2018})}\BibitemShut {NoStop}%
\bibitem [{\citenamefont {Schlipf}\ \emph {et~al.}(2018)\citenamefont
  {Schlipf}, \citenamefont {Ponc\'e},\ and\ \citenamefont
  {Giustino}}]{Schlipf2018}%
  \BibitemOpen
  \bibfield  {author} {\bibinfo {author} {\bibfnamefont {M.}~\bibnamefont
  {Schlipf}}, \bibinfo {author} {\bibfnamefont {S.}~\bibnamefont {Ponc\'e}}, \
  and\ \bibinfo {author} {\bibfnamefont {F.}~\bibnamefont {Giustino}},\ }\href
  {\doibase 10.1103/PhysRevLett.121.086402} {\bibfield  {journal} {\bibinfo
  {journal} {Phys. Rev. Lett.}\ }\textbf {\bibinfo {volume} {121}},\ \bibinfo
  {pages} {086402} (\bibinfo {year} {2018})}\BibitemShut {NoStop}%
\bibitem [{\citenamefont {Thouin}\ \emph {et~al.}(2018)\citenamefont {Thouin},
  \citenamefont {Neutzner}, \citenamefont {Cortecchia}, \citenamefont
  {Dragomir}, \citenamefont {Soci}, \citenamefont {Salim}, \citenamefont {Lam},
  \citenamefont {Leonelli}, \citenamefont {Petrozza}, \citenamefont {Kandada},\
  and\ \citenamefont {Silva}}]{Thouin2018}%
  \BibitemOpen
  \bibfield  {author} {\bibinfo {author} {\bibfnamefont {F.}~\bibnamefont
  {Thouin}}, \bibinfo {author} {\bibfnamefont {S.}~\bibnamefont {Neutzner}},
  \bibinfo {author} {\bibfnamefont {D.}~\bibnamefont {Cortecchia}}, \bibinfo
  {author} {\bibfnamefont {V.~A.}\ \bibnamefont {Dragomir}}, \bibinfo {author}
  {\bibfnamefont {C.}~\bibnamefont {Soci}}, \bibinfo {author} {\bibfnamefont
  {T.}~\bibnamefont {Salim}}, \bibinfo {author} {\bibfnamefont {Y.~M.}\
  \bibnamefont {Lam}}, \bibinfo {author} {\bibfnamefont {R.}~\bibnamefont
  {Leonelli}}, \bibinfo {author} {\bibfnamefont {A.}~\bibnamefont {Petrozza}},
  \bibinfo {author} {\bibfnamefont {A.~R.~S.}\ \bibnamefont {Kandada}}, \ and\
  \bibinfo {author} {\bibfnamefont {C.}~\bibnamefont {Silva}},\ }\href
  {\doibase 10.1103/PhysRevMaterials.2.034001} {\bibfield  {journal} {\bibinfo
  {journal} {Phys. Rev. Materials}\ }\textbf {\bibinfo {volume} {2}},\ \bibinfo
  {pages} {034001} (\bibinfo {year} {2018})}\BibitemShut {NoStop}%
\bibitem [{\citenamefont {Thouin}\ \emph {et~al.}(2019)\citenamefont {Thouin},
  \citenamefont {Cortecchia}, \citenamefont {Petrozza}, \citenamefont
  {Srimath~Kandada},\ and\ \citenamefont {Silva}}]{Thouin2019PRR}%
  \BibitemOpen
  \bibfield  {author} {\bibinfo {author} {\bibfnamefont {F.}~\bibnamefont
  {Thouin}}, \bibinfo {author} {\bibfnamefont {D.}~\bibnamefont {Cortecchia}},
  \bibinfo {author} {\bibfnamefont {A.}~\bibnamefont {Petrozza}}, \bibinfo
  {author} {\bibfnamefont {A.~R.}\ \bibnamefont {Srimath~Kandada}}, \ and\
  \bibinfo {author} {\bibfnamefont {C.}~\bibnamefont {Silva}},\ }\href
  {\doibase 10.1103/PhysRevResearch.1.032032} {\bibfield  {journal} {\bibinfo
  {journal} {Phys. Rev. Research}\ }\textbf {\bibinfo {volume} {1}},\ \bibinfo
  {pages} {032032} (\bibinfo {year} {2019})}\BibitemShut {NoStop}%
\end{thebibliography}
%\end{document}
\end{document}